\documentclass[11pt,a4paper]{article}

\usepackage[T1]{fontenc}
\usepackage[utf8]{inputenc}
\usepackage[margin=2.5cm]{geometry}
\usepackage{amsmath,amssymb,amsthm}
\usepackage{mathtools}
\usepackage{graphicx}
\usepackage{booktabs}
\usepackage{array}
\usepackage{tabularx}
\usepackage{enumitem}
\usepackage{xcolor}
\usepackage[hidelinks]{hyperref}
\usepackage[authoryear,round]{natbib}
\usepackage{microtype}
\usepackage{setspace}

\theoremstyle{definition}
\newtheorem{definition}{Definition}[section]
\newtheorem{remark}[definition]{Remark}
\theoremstyle{plain}
\newtheorem{proposition}{Proposition}[section]

\newcommand{\R}{\mathbb{R}}

\title{%
Learning Latent Memory States from Longitudinal Athlete Monitoring Data%
}

\author{%
Dae-Jin Lee%
\thanks{IE University, School of Science and Technology,
Cardenal Zu\~{n}iga, S/N, 40003 Segovia, Spain.
Email: \texttt{Dae-Jin.Lee@ie.edu}.
ORCID: \texttt{0000-0002-8995-8535}.}%
}

\date{6 August 2026}

\begin{document}
\maketitle

\begin{abstract}
We propose a new unit of analysis for longitudinal data: the Latent Memory Table.
The scientific contribution is not the encoder. It is that table, treated as a reusable statistical object on the same footing as a matrix of principal-component scores, a table of estimated random effects, or a table of predicted probabilities.
We estimate a statistical table that summarizes recent longitudinal history and is intended to be stored, queried, analysed and reused throughout the statistical workflow.
A memory operator maps each masked windowed history to a finite-dimensional state; collecting those states with uncertainty yields the Latent Memory Table.
Validation is organized around six properties---recoverability, personalization, temporal coherence, interpretability, stability and reusability---summarized by a composite quality index \(Q\); the Transformer, the SoccerMon case study and the simulations exist to argue that this table deserves that status.
Classical exponentially weighted moving averages and related short- and long-horizon scalar summaries arise as restricted, typically univariate special cases of the same operator class.
A simulation study with known memory mechanisms shows that \(Q\) and rotation-invariant recovery scores discriminate genuine multivariate or personalized memory from negative controls and from misspecified windows, whereas regime classification accuracy alone does not.
SoccerMon serves as an empirical case study: a constructed Latent Memory Table attains \(Q\approx 0.73\) versus about \(0.40\) for classical and lagged principal-component baselines, with incremental held-out value for some wellness targets and Procrustes ensembles for row-wise reliability.
\end{abstract}



%==============================================================================
\section{Introduction}
\label{sec:intro}
%==============================================================================

\label{sec:problem}
Longitudinal monitoring increasingly records irregular, multivariate, partially observed histories for each unit---subjects, patients, athletes or devices---and the statistical task is to retain what matters from recent history as a validated statistical object that later analyses can reuse.
A common modelling reflex is to compress recent history into a few \emph{prespecified} scalar summaries---for example exponentially weighted moving averages (EWMA) or other fixed-horizon smoothers---or to train end-to-end predictors of an application-dependent endpoint.
Prespecified scalars are typically univariate, discard cross-variable interactions and missingness structure, and cannot adapt to the observed multivariate dependence.
End-to-end predictors optimize short-term scores while leaving underspecified the scientific object of interest: the validated Latent Memory Table.

Elite athlete monitoring is a concrete instance of this problem: subjective self-reports (session rating of perceived exertion, wellness) mix with sparse objective Global Positioning System (GPS) and heart-rate signals
\citep{gabbett2016,hulin2016}.
We use it as a \emph{case study}, not as the boundary of the methodology.

\label{sec:proposal}
We reframe the problem around a \emph{memory operator}
\[
\mathcal{M}_\theta:\mathcal{H}_W\to\R^{d},
\qquad
m_{i}(t)=\mathcal{M}_\theta(H_{i,t}),
\]
mapping a masked history \(H_{i,t}\) over a window of length \(W\) to a fixed-dimensional \emph{latent memory state} \(m_{i}(t)\).
Collecting states with uncertainty yields the \emph{Latent Memory Table}
\(\mathcal{T}=\{\widehat m_i(t),\widehat\Sigma_i(t)\}\)
(Figure~\ref{fig:product}): a rectangular object with columns for unit, date, coordinates and uncertainty.
\begin{quote}
\emph{We propose a new unit of analysis for longitudinal data: the Latent Memory Table.}\\[0.35em]
\emph{The scientific contribution is not the encoder. It is the validated Latent Memory Table.}\\[0.35em]
\emph{We estimate a statistical table that summarizes recent longitudinal history.}
\end{quote}
Unlike a prediction score or a transient embedding, a Latent Memory Table is intended to be stored, queried, analysed and reused throughout the statistical workflow---on the same footing as a matrix of principal-component scores, estimated random effects, or a table of predicted probabilities.
Any sequence encoder (Transformer, long short-term memory network (LSTM), linear map, PCA of lags) is only an estimator \(\widehat{\mathcal{M}}_\theta\) of the memory operator.
Classical EWMA-type smoothers belong to the same operator class as low-dimensional special cases (Section~\ref{sec:classical}).
The Transformer, SoccerMon and the property system \(\mathcal{P}\) are instruments in service of that claim: a \(\mathcal{P}\)-validated Latent Memory Table supports statistical operators---paths, anomaly scores, neighbourhoods, clustering and covariate use---without retraining (Section~\ref{sec:applied}; property \(P_6\)), and can add information beyond classical fixed scalar summaries when those are available.
The central question is therefore not whether a network classifies well, but whether the Latent Memory Table deserves to be treated as a reusable unit of analysis, under known simulation mechanisms and on an empirical case study
(Sections~\ref{sec:desiderata}, \ref{sec:sim}, \ref{sec:incremental}).
\label{sec:scope}
We formalize the memory operator, treat classical fixed scalar summaries as special cases, and define usefulness via a property system \(\mathcal{P}=\{P_1,\dots,P_6\}\) and a composite quality index \(Q\) (Section~\ref{sec:desiderata}).
The simulation study asks whether a constructed Latent Memory Table recovers known memory mechanisms---and fails negative controls---on the same \(\mathcal{P}/Q\) scale used empirically; SoccerMon is then an empirical \emph{case study} of multivariate longitudinal athlete monitoring, including an incremental-value check against classical workload scalars when those are present.
We do not claim encoder uniqueness, universal optimality of \(W=28\), physiological truth, causality or endpoint prediction.
Models fitted later on rows of the Latent Memory Table---mixed-effects and other regression models, survival or joint models, functional data analyses, time-series forecasting, clustering---are licensed by \(P_6\) but are not primary analyses here; Section~\ref{sec:applied} only maps that operator family.
Software is outlined in Section~\ref{sec:software}.

%==============================================================================
\section{Background and related work}
\label{sec:related}
%==============================================================================

Irregular multivariate panels with informative missingness arise in clinical monitoring and in elite sport
\citep{soccermon2021,rossi2018,che2018}.
Domain practice often collapses recent history to EWMA-type scalars or rolling ratios with predetermined decay
\citep{gabbett2016,hulin2016}; those tools remain useful but are not a general memory representation.
To our knowledge there is no published multivariate Latent Memory Table for athlete monitoring that is validated independently of a single endpoint.

Medical informatics has developed encoders for sparse, irregular longitudinal data, including recurrent models with missingness mechanisms (e.g.\ GRU-D, \citealt{che2018}),
multi-time attention networks
\citep{shukla2021mtand}
and self-supervised Transformers for sparse clinical series
\citep{tipirneni2022}.
Recent lines push the same machinery further toward endpoint tasks:
Transformer-augmented latent-variable survival models
\citep{ogretir2025seqrisk},
Transformer--VAE clustering of longitudinal electronic health records jointly with survival
\citep{qiu2025vadesc}
and continuous-time longitudinal representation learning for disease progression
\citep{zeghlache2024latim}.
In each case the product is a predictor (or a task-tied latent) rather than an exported, multi-purpose table.
Closer in spirit, but still distinct, are latent spline models for intensively collected ordinal digital-health series
\citep{lunt2025splines}
and classical alignment/prediction pipelines for irregular repeated measurements
\citep{eekhout2023}.
Those approaches address irregular longitudinal structure without framing classical scalar summaries as restricted memory operators, without a multi-property validation system for an exported Latent Memory Table, and without separating estimation of that table from endpoint modelling.

Mixed-effects, joint longitudinal--survival and functional data models provide the natural setting in which a validated Latent Memory Table \(\mathcal{T}\) can later enter as a time-varying covariate or functional input
\citep{tsiatis2004,rizopoulos2012,ramsay2005}.
We construct and \(\mathcal{P}\)-validate \(\mathcal{T}\) so those operators have a well-defined input; fitting a particular endpoint model is left as subsequent work enabled by the table.
Masked sequence encoders for irregular panels are therefore well explored in computational medicine, whereas the present reframing---the Latent Memory Table as reusable statistical object, with classical summaries as special cases, property system \(\mathcal{P}\) with composite \(Q\), table-level uncertainty, and mechanism-recovery simulation with negative control---is the methodological gap we address.

%==============================================================================
\section{Latent memory states and a property system for validation}
\label{sec:framework}
%==============================================================================

This section formalizes the Latent Memory Table motivated in Section~\ref{sec:intro}.
We introduce a memory operator that maps a masked windowed history to a latent memory state, collect states into \(\mathcal{T}\), place classical scalar summaries as restricted special cases of the same operator class, and state a property system \(\mathcal{P}\) under which usefulness of \(\mathcal{T}\) is assessed.
Latent coordinates are identified only up to rotation, so subsequent inference focuses on geometry and validated functionals rather than on named axes.

\label{sec:definition}
Let \(i\in\{1,\dots,N\}\) index athletes and \(t\) a discrete day.
For window length \(W\), the masked history
\[
H_{i,t}
=
\bigl\{(\mathbf{x}_{i,s},\,\delta_{i,s})\bigr\}_{s=t-W+1}^{t}
\in\mathcal{H}_W
\]
collects channel values \(\mathbf{x}_{i,s}\) and missingness masks \(\delta_{i,s}\).

\begin{definition}[Memory operator and latent memory state]
\label{def:lms}
A \emph{memory operator} is a map
\begin{equation}
\label{eq:op}
\mathcal{M}_\theta:\mathcal{H}_W\to\R^{d}.
\end{equation}
The associated \emph{latent memory state} is
\begin{equation}
\label{eq:m}
m_{i}(t)=\mathcal{M}_\theta(H_{i,t})\in\R^{d}.
\end{equation}
Given data, an encoder yields an estimated operator \(\widehat{\mathcal{M}}_\theta\) and
\begin{equation}
\label{eq:mhat}
\widehat m_{i}(t)=\widehat{\mathcal{M}}_\theta(H_{i,t}).
\end{equation}
\end{definition}

The estimand \(\mathcal{T}\) is defined relative to a history domain \(\mathcal{H}_W\), dimension \(d\), a training criterion (up to its invariances) and the property system \(\mathcal{P}\) (Section~\ref{sec:desiderata})---not as an arbitrary encoder feature map or an injury score.

\begin{remark}[Working summary, not formal sufficiency]
\label{rem:working}
We treat \(m_{i}(t)\) as a working history summary.
Formal sufficiency would require a fully specified likelihood and target functional
\citep{tishby2000,wainwright2008}; that claim is not made here.
\end{remark}

\begin{remark}[Rotation non-identifiability]
\label{rem:rotate}
If \(\widehat m\) satisfies the training criterion, so typically does \(Q\widehat m\) for orthogonal \(Q\).
Individual coordinates are therefore not intrinsically identified without constraints.
Inference should focus on rotation-invariant or post-processed objects: pairwise distances/cosines, neighbourhoods, subspaces, principal directions after training, variance decompositions and predictive information for auxiliary targets on the Latent Memory Table.
Across random seeds we compare geometry after orthogonal Procrustes alignment
\citep{gower1975}
rather than raw coordinates.
\end{remark}

\begin{definition}[Latent Memory Table]
\label{def:table}
The \emph{Latent Memory Table} is the collection
\begin{equation}
\label{eq:T}
\mathcal{T}
=
\Bigl\{\bigl(\widehat m_i(t),\,\widehat\Sigma_i(t)\bigr): i=1,\dots,N,\; t\in\mathcal{T}_i\Bigr\},
\end{equation}
where \(\widehat m_i(t)\) is a Procrustes-aligned ensemble mean state and \(\widehat\Sigma_i(t)\) is the corresponding empirical covariance across training replicates (Section~\ref{sec:uncertainty}).
Row-level reliability summaries (e.g.\ \(\operatorname{tr}\widehat\Sigma_i(t)\), ensemble cosine similarity, mask sensitivity) belong to \(\mathcal{T}\).
Methods are compared by the quality of the Latent Memory Table \(\mathcal{T}\), not by the encoder that produced it.
\end{definition}

\begin{figure}[t]
\centering
\includegraphics[width=\textwidth]{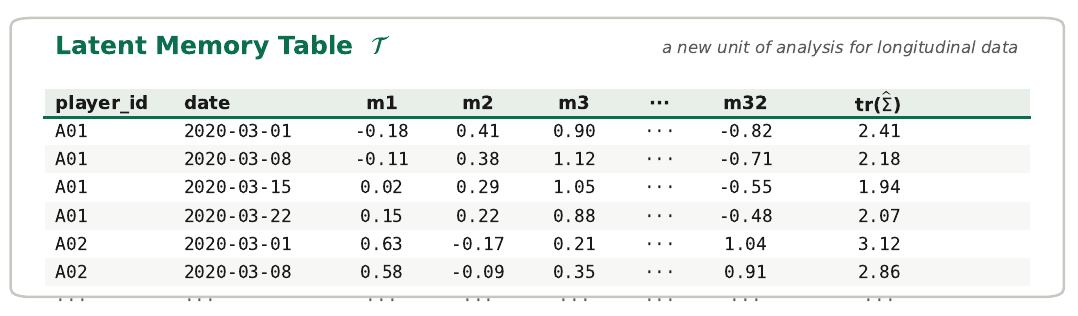}
\caption{Schematic excerpt of the exported Latent Memory Table \(\mathcal{T}\): one row per athlete--window with state coordinates and uncertainty.
We propose a new unit of analysis for longitudinal data: the Latent Memory Table---meant to be stored, queried, analysed and reused, not an embedding.}
\label{fig:product}
\end{figure}

Operationally, the algorithm returns a rectangular export: one row per athlete--window with identifiers, date, coordinates \(\widehat m_1,\dots,\widehat m_d\) and a reliability summary such as \(\operatorname{tr}(\widehat\Sigma_i(t))\)
(Figure~\ref{fig:product}).
That export is the scientific product: a persistent unit of analysis, not a disposable embedding.

\label{sec:classical}
Across domains, recent longitudinal exposure is often compressed into fixed scalar memory summaries.
The prototype is the EWMA of a univariate series \(x_t\),
\(z_t=\lambda x_t+(1-\lambda)z_{t-1}\)
with predetermined decay \(\lambda\in(0,1]\).
Short- and long-horizon rolling or EWMA summaries, and ratios of two such horizons, are instances of the same class
\citep{banister1991,foster2001,gabbett2016,hulin2016,murray2017,lolli2019,windt2019};
in the SoccerMon case study they appear as acute training load (ATL), chronic training load (CTL) and the acute:chronic workload ratio (ACWR).
We treat these objects as \emph{degenerate memory operators}: useful as fixed dashboards, yet typically univariate, misspecification-blind and missingness-agnostic relative to a joint history \(H_{i,t}\).
The table \(\mathcal{T}\) targets the complementary object---a multivariate, missingness-aware learned summary---of which fixed scalars are special cases (Proposition~\ref{prop:ewma-deg}).

\label{sec:desiderata}
Usefulness of \(\mathcal{T}\) is membership in a finite \emph{property system}
\begin{equation}
\label{eq:P}
\mathcal{P}
=
\{P_1,\dots,P_6\},
\end{equation}
each \(P_j\) a formal requirement on \(\mathcal{M}_\theta\) (equivalently, on \(\mathcal{T}\)).
Scores \(S_j(\mathcal{T})\in[0,1]\) are empirical \emph{tests} of those requirements, not the requirements themselves.

\begin{definition}[Property system \(\mathcal{P}\)]
\label{def:P}
Let \(m_{i}(t)=\mathcal{M}_\theta(H_{i,t})\) and let \(\mathcal{R}\) denote a known longitudinal regime label (here Team\(\times\)Season). Write \(S\) for a similarity on \(\R^{d}\) (cosine unless stated otherwise). The properties are:
\begin{enumerate}[leftmargin=1.6em,itemsep=0.35em,label=\(P_{\arabic*}\).]
\item \emph{Recoverability.}
There exists a (possibly randomized) decoder \(\psi\) such that \(\psi(m_{i}(t))\) recovers \(\mathcal{R}_{i,t}\) above chance; equivalently, \(I(m_{i}(t);\mathcal{R}_{i,t})>0\) at a nontrivial rate.
\item \emph{Personalization.}
Within-athlete similarity dominates cross-athlete similarity,
\[
\mathbb{E}\bigl[S(m_{i}(t),m_{i}(t'))\bigr]
>
\mathbb{E}\bigl[S(m_{i}(t),m_{j}(s))\bigr],
\qquad i\neq j,
\]
after removing trivial adjacency induced by overlapping windows.
\item \emph{Temporal coherence.}
The lag profile
\[
\rho_m(h)
=
\mathbb{E}\bigl[\cos\{m_{i}(t),m_{i}(t+h)\}\bigr]
\]
is nonincreasing in \(h\) on the scale of the memory horizon, absent abrupt process changes.
\item \emph{Interpretability (alignment).}
There exist low-dimensional summaries \(u(m_{i}(t))\) (e.g.\ leading PCs) and interpretable constructs \(x_{i}(t)\) (e.g.\ exposure or response channels) such that conditional independence fails:
\(u(m_{i}(t))\not\!\perp\!\!\!\perp x_{i}(t)\mid\mathcal{R}\)
(equivalently, residual association remains after conditioning on regime).
\item \emph{Stability.}
The geometry of \(\mathcal{T}\) (pairwise similarities and recoverable subspaces) varies continuously in the operating choices \((W,d,\mathrm{stride},\mathrm{mask},\mathrm{encoder})\) on a prespecified neighbourhood of the reported configuration.
\item \emph{Reusability.}
For a family of targets \(\{Y^{(1)},\dots,Y^{(K)}\}\) not used to train \(\mathcal{M}_\theta\), the fixed table retains incremental predictive information beyond a classical baseline for more than one \(Y^{(k)}\).
\end{enumerate}
\end{definition}

\begin{definition}[\(\mathcal{P}\)-validity]
\label{def:valid}
A table \(\mathcal{T}\) is \emph{\(\mathcal{P}\)-valid} if it satisfies \(P_1,\dots,P_6\) in Definition~\ref{def:P}, as assessed by the tests in Table~\ref{tab:desiderata} and, where available, by simulation recovery of a known memory process.
\end{definition}

Normalized scores \(S_j(\mathcal{T})\in[0,1]\) operationalize each test.
The \emph{Latent Memory Table Quality}
\begin{equation}
\label{eq:Q}
Q(\mathcal{T})
=
\frac1{6}\sum_{j=1}^{6} S_j(\mathcal{T})
\end{equation}
is an equally weighted aggregate test statistic for \(\mathcal{P}\).
Alternative weights are sensitivity analyses.
\(Q\) compares \emph{tables} (learned, classical, PCA,\ldots), not architectures in isolation.

\begin{proposition}[Classical summaries as degenerate cases of \(\mathcal{P}\)]
\label{prop:ewma-deg}
Fix a univariate load series \(x_t\) and \(\lambda\in(0,1]\).
The EWMA recursion \(z_t=\lambda x_t+(1-\lambda)z_{t-1}\) induces a memory operator
\(\mathcal{M}^{\mathrm{EWMA}}_{\lambda}:\mathcal{H}_W\to\R\)
with \(d=1\), no cross-channel mixing and no mask dependence.
Windowed or dual-horizon EWMA summaries, and scalar ratios of two such memories, are instances of the same class.
Relative to \(\mathcal{P}\), these operators are \emph{degenerate}:
\begin{itemize}[leftmargin=1.4em,itemsep=0.15em]
\item \(P_1\) can hold only for regimes visible in the single channel \(x\);
\item \(P_2\) reduces to path identity of a shared scalar smoother (no learned subject-specific geometry);
\item \(P_4\)--\(P_6\) cannot exploit multivariate missingness-aware history beyond what is already collapsed into \(z_t\) (or a short/long pair);
\item \(P_5\) is exhausted by sensitivity in \(\lambda\) (or the two window lengths), not by a representation dimension or encoder class.
\end{itemize}
They sit inside the operator formalism but occupy a low-dimensional corner of \(\mathcal{P}\).
\end{proposition}

\begin{table}[t]
\centering
\small
\caption{Property system \(\mathcal{P}\) and corresponding empirical tests.}
\label{tab:desiderata}
\renewcommand{\arraystretch}{1.25}
\begin{tabular}{@{}
  >{\raggedright\arraybackslash}p{0.18\linewidth}
  >{\raggedright\arraybackslash}p{0.38\linewidth}
  >{\raggedright\arraybackslash}p{0.36\linewidth}
  @{}}
\toprule
\textbf{Property} & \textbf{Formal content} & \textbf{Empirical test} \\
\midrule
\(P_1\) Recoverability
  & Regime information retained in \(m_{i}(t)\)
  & Regime classification; subspace recovery in simulations \\
\addlinespace
\(P_2\) Personalization
  & Within-athlete \(>\) between-athlete similarity
  & Cosine contrasts; subject-identity classification; sim calibration \\
\addlinespace
\(P_3\) Temporal coherence
  & Gradual lag decay of \(\rho_m(h)\)
  & Lagged cosine; persistence recovery in simulations \\
\addlinespace
\(P_4\) Interpretability
  & Residual alignment with domain constructs given \(\mathcal{R}\)
  & Conditional associations with domain constructs \\
\addlinespace
\(P_5\) Stability
  & Geometry stable in \((W,d,\mathrm{stride},\ldots)\)
  & Sensitivity grid (partially reported) \\
\addlinespace
\(P_6\) Reusability
  & Multi-target information beyond classical baselines
  & Fixed-representation multi-outcome prediction \\
\bottomrule
\end{tabular}
\end{table}

%==============================================================================
\section{Estimation}
\label{sec:estimation}
%==============================================================================

The encoder below is only the computational means of constructing \(\mathcal{T}\).

\label{sec:encoder}
Each day is represented by the concatenation of standardized channel values and their availability masks, together with sinusoidal positional encoding.
We use a compact two-layer Transformer encoder with mean pooling over observed days (padding masked) to produce \(m_{i}(t)\in\R^{d}\).
Operating choices are \(W=28\) (chronic-load horizon), stride~7, twelve value channels plus twelve masks, and \(d=32\); full hyperparameters appear in Appendix~\ref{app:arch}.
Sensitivity to \(d\in\{16,32,64\}\) is in Appendix~\ref{app:repro}.
Training uses AdamW
\citep{loshchilov2019}
with the auxiliary Team\(\times\)Season criterion of Section~\ref{sec:objective}; checkpoints are selected by held-out regime accuracy.

\label{sec:objective}
We train with an \emph{auxiliary} Team\(\times\)Season classification criterion: each window is labelled by the crossed factor of club and season, yielding four classes
\{Team~A--2020, Team~A--2021, Team~B--2020, Team~B--2021\},
optimized by weighted cross-entropy.
The criterion is not the estimand: it only forces rows of \(\mathcal{T}\) to retain regime information for testing \(P_1\).
Richer criteria (reconstruction, contrastive personalization) are left for future work.

\label{sec:windows}
Train/test splitting is \emph{player-grouped}: all windows of held-out athletes form the test set, so evaluation does not leak identity through overlapping windows of the same player.
In the reported run: 50 training athletes (4{,}216 windows) and 16 test athletes (1{,}261 windows), using the \(W=28\), stride-7 scheme of Appendix~\ref{app:arch}.

\begin{remark}[Standardization]
\label{rem:std}
Channel standardization is fit on training athletes only and then applied to all windows (training and held-out), preventing leakage of test-athlete scale information into the estimator.
\end{remark}

%==============================================================================
\section{Empirical case study: SoccerMon}
\label{sec:data}
%==============================================================================

\label{sec:source}
As an empirical case study of the methodology---not as its definitional domain---we analyse anonymized monitoring from two professional football teams (Team~A, Team~B) across seasons 2020 and 2021
\citep{soccermon2021}.
SoccerMon combines two complementary data layers:
\begin{itemize}[leftmargin=1.4em,itemsep=0.15em]
\item Subjective / self-reported streams
(from the club wellness and training-load questionnaires):
session-RPE and derived internal load, ATL, CTL and ACWR, together with daily wellness items (fatigue, soreness, sleep quality and duration, stress, mood, readiness), plus match-exposure, illness and injury event logs;
\item Objective / sensor streams:
sparse GPS and heart-rate summaries recorded on instrumented sessions.
\end{itemize}
The operator is estimated on the joint multivariate history, so rows of \(\mathcal{T}\) mix self-report and device channels under a common missingness mask.

\begin{remark}[Subjective measurement]
\label{rem:subjective}
Wellness and RPE are athlete self-reports with discrete scales, possible reporting bias and intermittent non-response.
Associations between principal directions of \(\widehat m_i(t)\) and sleep, fatigue or readiness should therefore be read as alignment with \emph{reported} constructs, not as validation against laboratory physiology.
\end{remark}

\label{sec:coverage}
Missingness is severe and systematic, and differs by data layer: GPS and HR are often available only on instrumented sessions; subjective wellness is intermittently missing; internal-load coverage is higher but not complete
(Appendix~\ref{app:missing}, Figure~\ref{fig:missing}).
Masks \(\delta_{i,s}\) are therefore part of \(H_{i,t}\).

\label{sec:window-table}
Calendar-aligned 28-day windows with stride 7 yield 5{,}477 windows from 66 athletes with sufficient history (Table~\ref{tab:windows}).

\begin{table}[t]
\centering
\caption{Windows and athletes by Team\(\times\)Season.}
\label{tab:windows}
\begin{tabular}{@{}lrrr@{}}
\toprule
Regime & Windows & Athletes & Mean observed days / window \\
\midrule
TeamA--2020 & 1323 & 27 & 28.0 \\
TeamA--2021 & 1474 & 30 & 28.0 \\
TeamB--2020 & 1350 & 31 & 28.0 \\
TeamB--2021 & 1330 & 32 & 28.0 \\
\midrule
Total & 5477 & 66 & --- \\
\bottomrule
\end{tabular}
\end{table}

%==============================================================================
\section{Validating the Latent Memory Table}
\label{sec:ladder}
%==============================================================================

Each row of \(\mathcal{T}\) is an athlete--time pair \((i,t)\) with coordinates \(\widehat m_i(t)\in\R^{d}\).
Because axes may rotate (Remark~\ref{rem:rotate}), we validate table \emph{geometry}---distances, neighbourhoods, trajectories, subspaces and utility---against \(\mathcal{P}\) (Definition~\ref{def:P}), not raw column names.
Team\(\times\)Season classification is only the auxiliary criterion used to estimate the operator.
The subsections below follow the ladder \(P_1\)--\(P_4\).

\begin{figure}[t]
\centering
\includegraphics[width=0.82\textwidth]{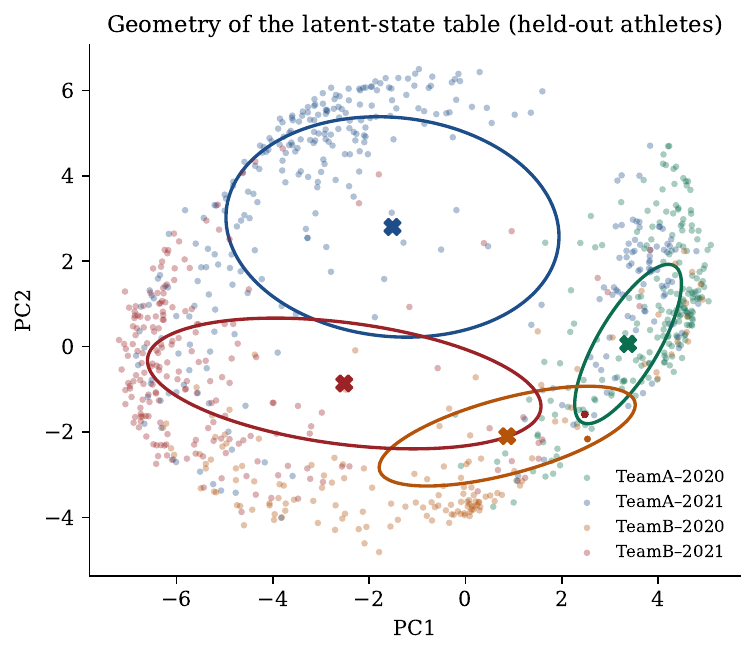}
\caption{Geometry of the table on held-out athletes: leading principal components (PC1--PC2) coloured by Team\(\times\)Season, with centroids and covariance ellipses.}
\label{fig:pca}
\end{figure}

\paragraph{Structure (\(P_1\)).}
\label{sec:ladder1}
Twenty-eight-day histories should produce states with recoverable systematic structure.
On held-out athletes, regime recovery reaches Transformer accuracy \(0.632\), linear classifier accuracy on fixed \(m\) \(0.645\) and \(k\)NN purity \(0.651\), against chance \(0.25\)
(Table~\ref{tab:ladder1}).
A linear classifier on fixed \(m\) matching (or exceeding) the neural classification head shows that the information resides in the table rows, not only in the head.
PCA of held-out states shows partial regime separation with overlap and regime centroids
(Figure~\ref{fig:pca}); silhouette \(0.090\) confirms that regimes are not cleanly clustered---as expected under shared sport physiology.

\begin{table}[t]
\centering
\caption{Structure of the Latent Memory Table (Team\(\times\)Season). Chance \(=0.25\).}
\label{tab:ladder1}
\begin{tabular}{@{}lc@{}}
\toprule
Metric & Value \\
\midrule
Transformer test accuracy & 0.632 \\
Linear classifier on fixed \(m\) (player-grouped) & 0.645 \\
\(k\)NN regime purity (\(k=15\)) & 0.651 \\
Silhouette (Team\(\times\)Season) & 0.090 \\
\bottomrule
\end{tabular}
\end{table}

\begin{figure}[t]
\centering
\includegraphics[width=0.78\textwidth]{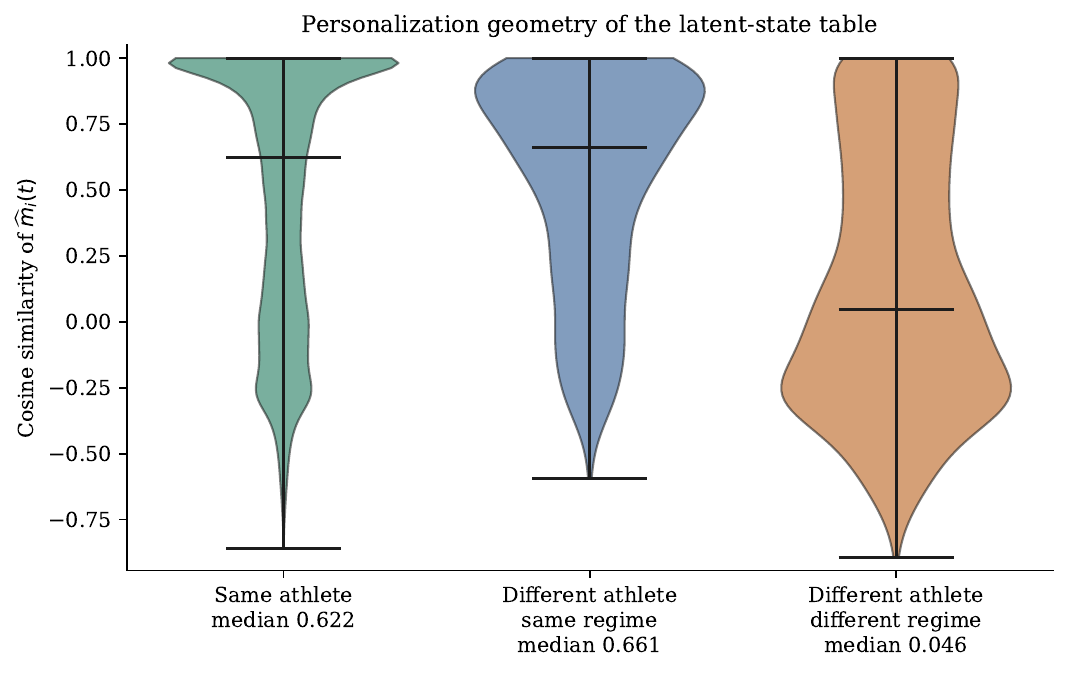}
\caption{Three-way personalization geometry.
Same-athlete pairs exclude adjacent windows (minimum lag of two steps).}
\label{fig:within}
\end{figure}

\paragraph{Personalization beyond regime (\(P_2\)).}
\label{sec:ladder2}
The table should not be only a Team\(\times\)Season code; it should retain athlete-level heterogeneity.
Coordinates are not interpreted individually; we compare pairwise geometry.
A three-way cosine comparison separates (i)~same athlete, (ii)~different athlete / same regime and (iii)~different athlete / different regime, using pairs at least two window steps apart to reduce mechanical overlap from stride-7 windows
(Figure~\ref{fig:within}).
Medians are approximately \(0.62\) (same athlete), \(0.66\) (same regime) and \(0.05\) (different regime).
Cross-regime separation is therefore the dominant pairwise signal under the present objective.
Same-athlete similarity remains far above cross-regime similarity, and subject-identity classification with a linear model on fixed states reaches accuracy \(0.241\) versus chance \(\approx 0.016\)
(Table~\ref{tab:ladder2}), so individual signal is present---but largely subordinate to regime geometry in pairwise cosine space.
This motivates richer personalization objectives and within-regime analyses below.

\begin{table}[t]
\centering
\caption{Personalization and temporal coherence of table rows.}
\label{tab:ladder2}
\begin{tabular}{@{}lc@{}}
\toprule
Metric & Value \\
\midrule
Same-athlete cosine (median; lag \(\ge 2\) steps) & 0.622 \\
Different-athlete, same-regime cosine & 0.661 \\
Different-athlete, different-regime cosine & 0.046 \\
Subject-identity linear classification accuracy & 0.241 \\
Subject-identity chance & 0.016 \\
Adjacent-window cosine (median) & 0.950 \\
Lag-3 window cosine (median) & 0.881 \\
Persistence horizon (first lag with median \(<0.90\)) & 9 steps \\
\bottomrule
\end{tabular}
\end{table}

\begin{figure}[t]
\centering
\includegraphics[width=0.925\textwidth]{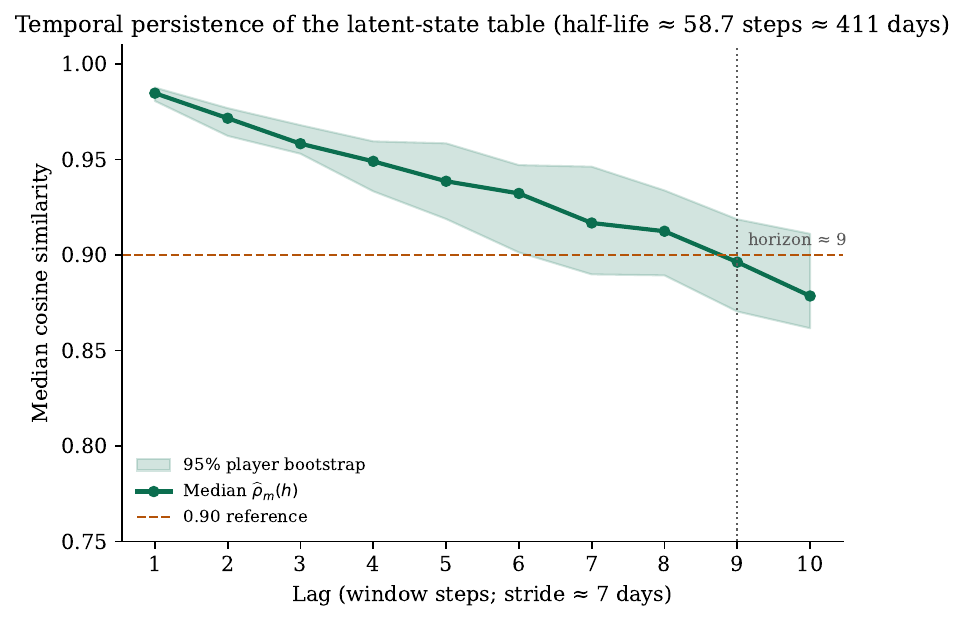}
\caption{Temporal persistence \(\widehat\rho_m(h)\) with 95\% player-bootstrap bands and a \(0.90\) reference line.}
\label{fig:acf}
\end{figure}

\begin{figure}[t]
\centering
\includegraphics[width=0.95\textwidth]{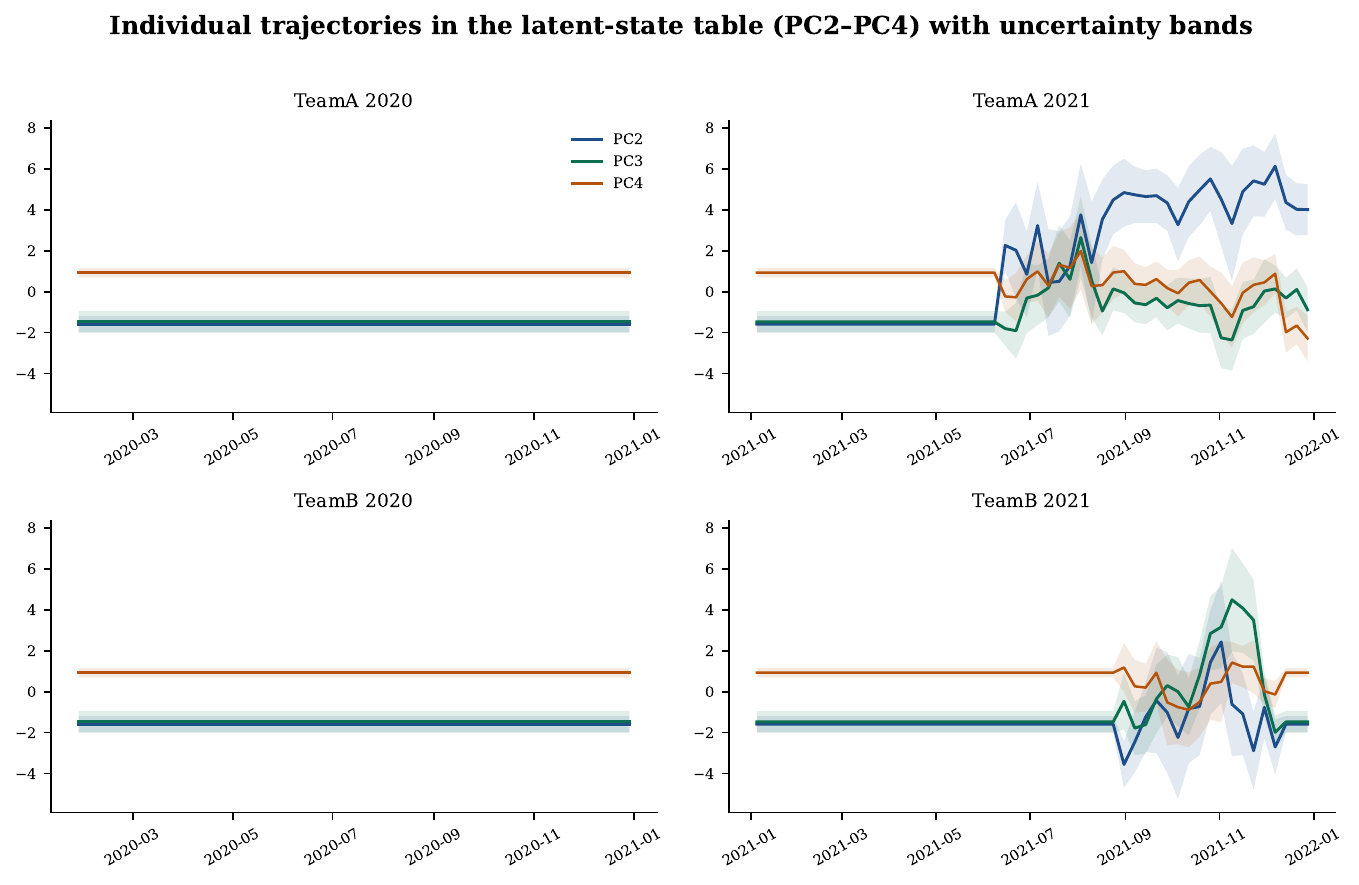}
\caption{Latent trajectories in PC2--PC4 for representative athletes from each Team\(\times\)Season, with ensemble uncertainty bands (\(B=50\) seeds).
Panels use a common y-scale; visually flatter PC curves therefore indicate lower within-panel temporal variation (relative stability), not a plotting artifact.}
\label{fig:traj}
\end{figure}

\paragraph{Temporal coherence (\(P_3\)).}
\label{sec:persistence}
\(\widehat m_i(t)\) should evolve as a temporally coherent state, not as an independent encoding of each window.
The persistence function
\[
\widehat\rho_m(h)
=
\operatorname{median}_{i,t}\cos\{\widehat m_i(t),\widehat m_i(t+h)\}
\]
decays smoothly with lag
(Figure~\ref{fig:acf}).
Player-level bootstrap bands remain tight; the first lag at which the median falls below \(0.90\) is \(h=9\) window steps (about nine weeks under stride~7).
Adjacent and lag-3 medians are \(0.950\) and \(0.881\).
Part of adjacent similarity is mechanical (overlapping days); the multi-lag decay and bootstrap uncertainty are the scientifically relevant features.
Individual PC trajectories illustrate that rows form paths rather than an unstructured cloud
(Figure~\ref{fig:traj}).
The plot is descriptive of state dynamics, not a prediction-accuracy display.
Using the same y-scale across Team\(\times\)Season panels makes amplitude comparisons interpretable: flatter curves indicate lower within-panel temporal variation (relative stability), whereas larger oscillations indicate stronger temporal movement in that component.
Ensemble bands summarize estimation reliability along each path.

\begin{figure}[t]
\centering
\includegraphics[width=0.975\textwidth]{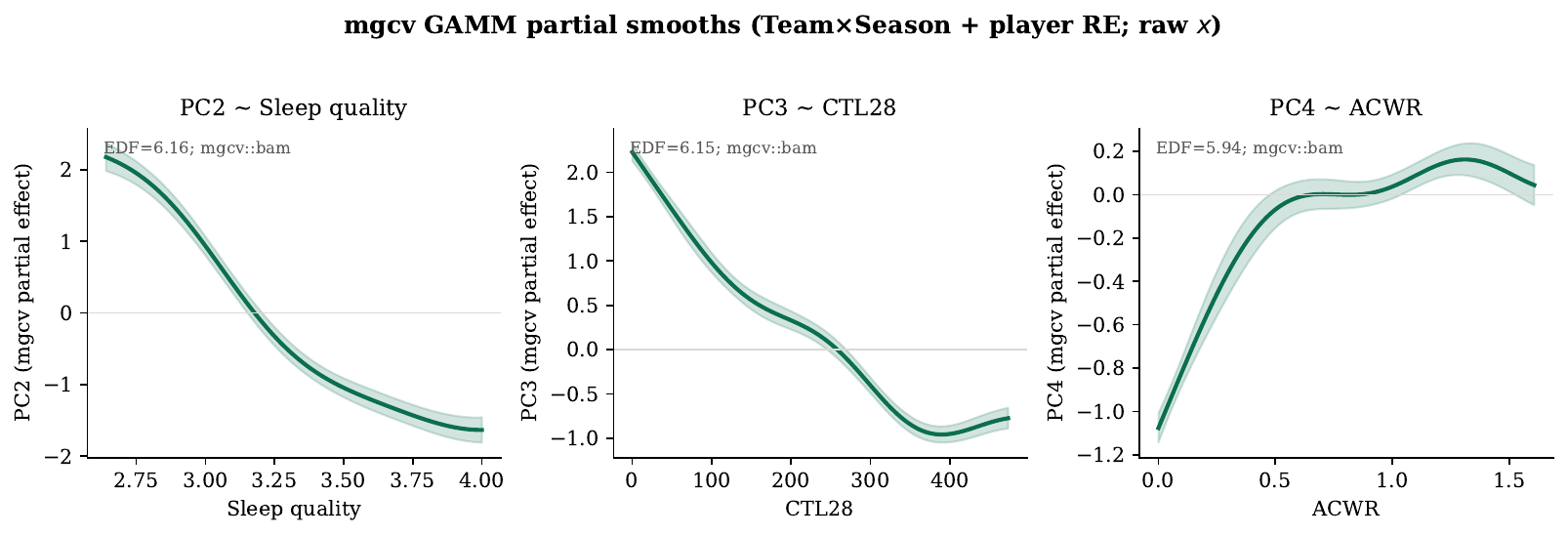}
\caption{\texttt{mgcv} GAMM partial smooths for PC2--sleep, PC3--CTL28 and PC4--ACWR
(Team\(\times\)Season fixed effects and player random intercept; bands are 95\% pointwise intervals for \(s(x)\); \(x\) on the raw physiological scale).}
\label{fig:gamm}
\end{figure}

\paragraph{Interpretability of subspaces (\(P_4\)).}
\label{sec:ladder3}
Principal directions of the table should align with \emph{reported} load and wellness constructs after conditioning on Team\(\times\)Season.
This is geometric alignment with self-report summaries, not physiological mechanism.
Leading associations include sleep with PC2 (\(r=0.558\); within-regime partial \(r=0.500\)) and CTL28/ATL/daily load with PC3 (\(r\approx 0.54\); partial \(r\approx 0.59\)--\(0.60\)); ACWR loads on PC4
(Table~\ref{tab:ladder3}).
The same associations are summarized as generalized additive mixed-model (GAMM) partial smooths fitted with the R package \texttt{mgcv}
\citep{wood2017}
(Team\(\times\)Season fixed effects and player random intercepts) in Figure~\ref{fig:gamm}.

\begin{table}[t]
\centering
\small
\caption{Selected PC--summary associations (Pearson \(r\) and within-regime partial \(r\)).}
\label{tab:ladder3}
\begin{tabular}{@{}llrrr@{}}
\toprule
PC & Summary & \(r\) & Partial \(r\) & \(n\) \\
\midrule
PC2 & Sleep quality & 0.558 & 0.500 & 3388 \\
PC3 & CTL28 & 0.546 & 0.604 & 5050 \\
PC3 & ATL & 0.543 & 0.599 & 5050 \\
PC3 & Daily load & 0.537 & 0.592 & 5050 \\
PC4 & ACWR & $-0.452$ & $-0.536$ & 5050 \\
PC2 & Stress & 0.353 & 0.306 & 3388 \\
PC2 & Fatigue & 0.287 & 0.238 & 3388 \\
PC2 & Soreness & 0.238 & 0.234 & 3388 \\
\bottomrule
\end{tabular}
\end{table}

%==============================================================================
\section{Statistical characterization of the table}
\label{sec:characterize}
%==============================================================================

Having checked \(P_1\)--\(P_4\) on geometry, we characterize how variance is shared across regimes and athletes, whether \(\mathcal{T}\) adds information beyond classical scalars (\(P_6\)), how tables score under the composite index \(Q\), and how uncertain the estimated states are (\(P_5\) under initialization).

\begin{figure}[t]
\centering
\includegraphics[width=0.92\textwidth]{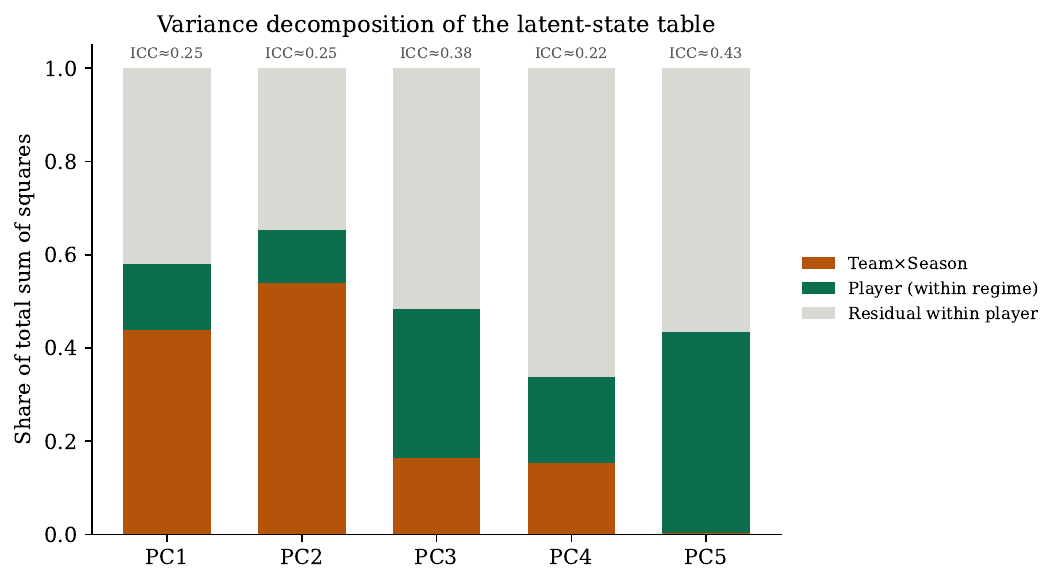}
\caption{Variance decomposition of PC1--PC5 of the Latent Memory Table.}
\label{fig:var}
\end{figure}

\paragraph{Variance decomposition.}
\label{sec:vc}
For the leading principal components (PC1--PC5) we decompose sums of squares into Team\(\times\)Season, player-within-regime and residual-within-player components
(Figure~\ref{fig:var}; Table~\ref{tab:vc}).
PC1--PC2 are largely regime-structured; PC3 and especially PC5 carry larger player shares.
Restricted maximum likelihood (REML) mixed models with Team\(\times\)Season fixed effects and player random intercepts yield intraclass correlation coefficients (ICCs) from about \(0.30\) (PC1--PC2) to \(0.47\) (PC3)
(Table~\ref{tab:vc}).
We treat Team and Season as fixed (two levels each), not as random populations.

\begin{table}[t]
\centering
\small
\caption{Variance shares (sum-of-squares) and REML player ICC for leading PCs.}
\label{tab:vc}
\begin{tabular}{@{}crrrrr@{}}
\toprule
PC & Regime share & Player share & Residual & ICC (REML) \\
\midrule
1 & 0.438 & 0.143 & 0.419 & 0.302 \\
2 & 0.539 & 0.114 & 0.348 & 0.301 \\
3 & 0.164 & 0.319 & 0.516 & 0.473 \\
4 & 0.153 & 0.185 & 0.662 & 0.363 \\
5 & 0.004 & 0.430 & 0.566 & 0.393 \\
\bottomrule
\end{tabular}
\end{table}

\begin{figure}[t]
\centering
\includegraphics[width=0.85\textwidth]{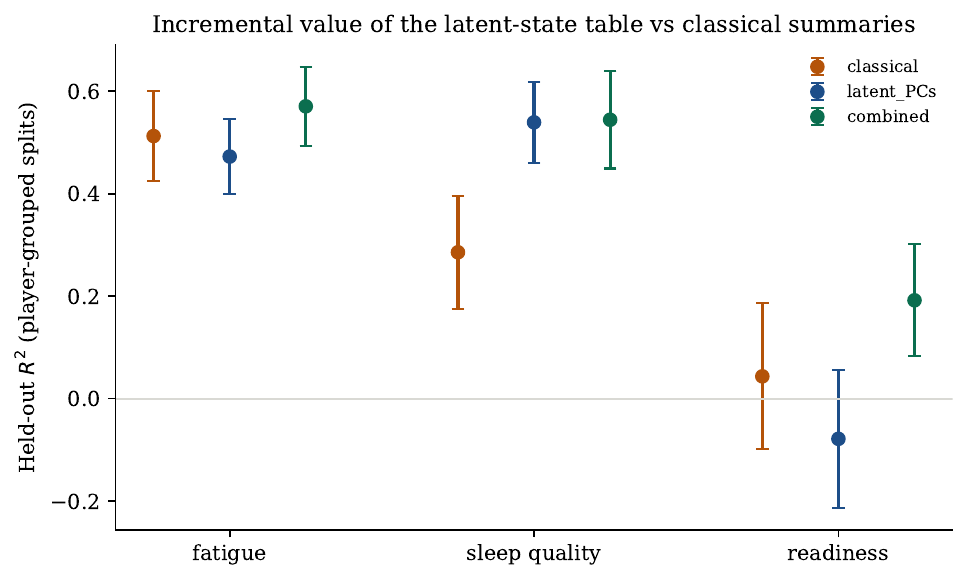}
\caption{Incremental value of the Latent Memory Table versus classical summaries for next-window wellness (mean \(\pm\) 1.96\,SE across player-grouped splits).}
\label{fig:incr}
\end{figure}

\paragraph{Incremental value versus classical summaries (\(P_6\)).}
\label{sec:incremental}
When classical fixed summaries are available in a given domain, a natural question is whether coordinates of \(\mathcal{T}\) add information beyond them.
As a reusability check on the SoccerMon case study---not as a primary prediction paper---we compare next-window fatigue, sleep quality and readiness under player-grouped holdout using (i)~classical load/wellness summaries excluding the outcome channel, (ii)~the first five PCs of \(\widehat m_i(t)\) and (iii)~both
(Figure~\ref{fig:incr}; Table~\ref{tab:incr}).
Combined models improve on classical alone for fatigue (\(R^2\) \(0.51\to 0.57\)) and readiness (\(0.04\to 0.19\)); for sleep quality, latent PCs alone already exceed classical (\(0.54\) vs \(0.29\)), with little further gain from combining.
This is case-study evidence that \(\mathcal{T}\) can add longitudinal information beyond fixed scalar workload summaries for some targets; it does not claim to replace those summaries, nor that every coordinate is needed in every model.

\begin{table}[t]
\centering
\small
\caption{Held-out \(R^2\) (player-grouped) for next-window wellness.}
\label{tab:incr}
\begin{tabular}{@{}lccc@{}}
\toprule
Outcome & Classical & Latent PCs & Combined \\
\midrule
Fatigue & 0.512 & 0.472 & 0.570 \\
Sleep quality & 0.286 & 0.539 & 0.544 \\
Readiness & 0.044 & $-0.078$ & 0.192 \\
\bottomrule
\end{tabular}
\end{table}

\begin{figure}[t]
\centering
\includegraphics[width=0.975\textwidth]{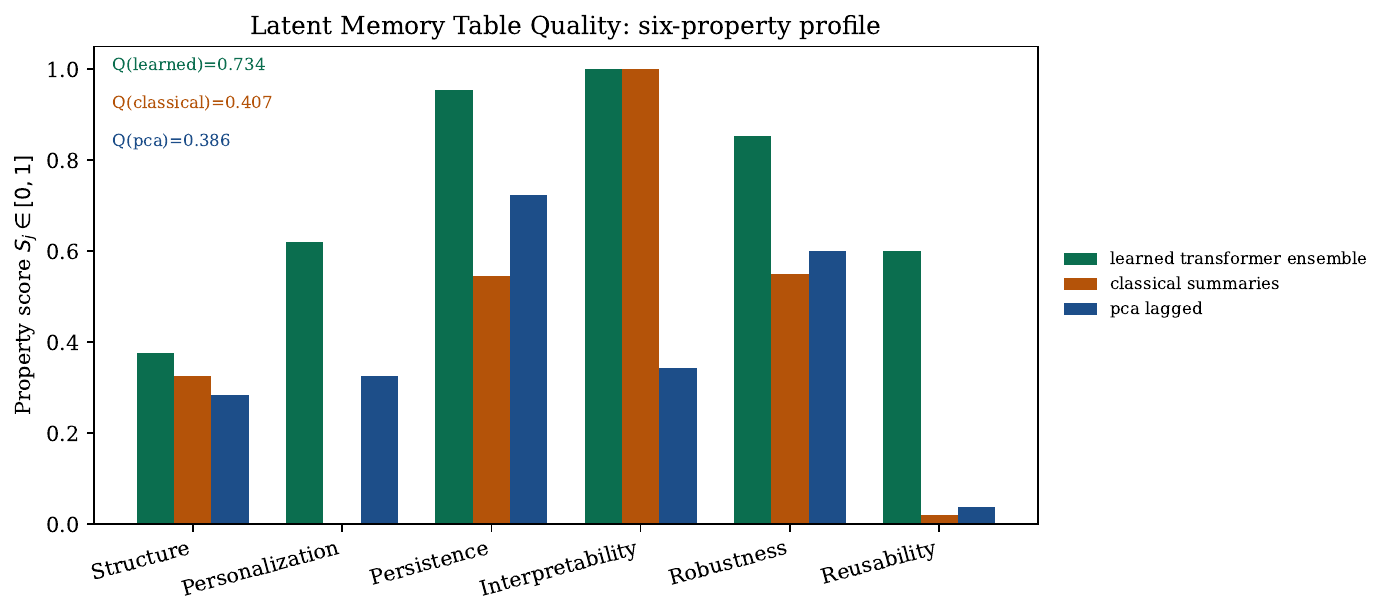}
\caption{Six-property quality profiles for learned, classical and PCA tables.}
\label{fig:Q}
\end{figure}

\paragraph{Latent Memory Table Quality.}
\label{sec:Q}
We score SoccerMon tables as tests of \(\mathcal{P}\)
(Figure~\ref{fig:Q}; Table~\ref{tab:Q}).
The learned ensemble table attains \(Q\approx 0.73\), versus \(\approx 0.41\) for concatenated classical summaries and \(\approx 0.39\) for lagged PCA.
Gains are concentrated in personalization, persistence, robustness (ensemble similarity) and reusability; interpretability scores for classical summaries are partly tautological because the same constructs enter both the table and the association library, so \(Q\) should be read as a profile, not a single oracle ranking.

\begin{table}[t]
\centering
\small
\caption{Latent Memory Table Quality on SoccerMon (normalized property scores \(S_j\) and composite \(Q\)).}
\label{tab:Q}
\begin{tabular}{@{}lccccccc@{}}
\toprule
Table & \(Q\) & Str. & Pers. & Persis. & Interp. & Rob. & Reuse \\
\midrule
Learned (ensemble) & 0.734 & 0.38 & 0.62 & 0.95 & 1.00 & 0.85 & 0.60 \\
Classical summaries & 0.407 & 0.33 & 0.00 & 0.55 & 1.00$^{*}$ & 0.55 & 0.02 \\
Lagged PCA & 0.386 & 0.28 & 0.33 & 0.72 & 0.34 & 0.60 & 0.04 \\
\bottomrule
\end{tabular}\\[0.3em]
{\footnotesize $^{*}$Classical interpretability is inflated by overlapping feature libraries.}
\end{table}

\paragraph{GAMM smooths and functional trajectories.}
\label{sec:gamm-fpca}
Beyond Pearson associations, the same \texttt{mgcv} GAMMs
\citep{wood2017}
with Team\(\times\)Season fixed effects and player random intercepts yield significant partial smooths for sleep\(\to\)PC2, CTL28\(\to\)PC3 and ACWR\(\to\)PC4
(Figure~\ref{fig:gamm}; Table~\ref{tab:gamm}; fitting details in Appendix~\ref{app:gamm}).
No causal interpretation is claimed.
Functional PCA of PC3 paths (Appendix~\ref{app:gamm}, Figure~\ref{fig:fpca}) shows that the leading FPC explains \(\approx 61\%\) of between-player curve variability, so rows of \(\mathcal{T}\) behave as functional longitudinal objects.

\begin{table}[t]
\centering
\small
\caption{GAMM summaries for \(\mathrm{PC}_{k,it}=s(x_{it})+\alpha_{g(i,t)}+b_i\)
(\(g\): Team\(\times\)Season regime; \(b_i\): player random intercept).}
\label{tab:gamm}
\begin{tabular}{@{}llrrrrr@{}}
\toprule
Response & Smooth \(x\) & \(n\) & EDF & \(p\) & \(R^2\) & \(\widehat\sigma_b\) \\
\midrule
PC2 & Sleep & 3388 & 6.16 & \(<10^{-16}\) & 0.817 & 1.15 \\
PC3 & CTL28 & 5050 & 6.15 & \(<10^{-16}\) & 0.610 & 0.61 \\
PC4 & ACWR & 5050 & 5.94 & \(<10^{-16}\) & 0.506 & 0.43 \\
\bottomrule
\end{tabular}
\end{table}

\begin{figure}[t]
\centering
\includegraphics[width=0.92\textwidth]{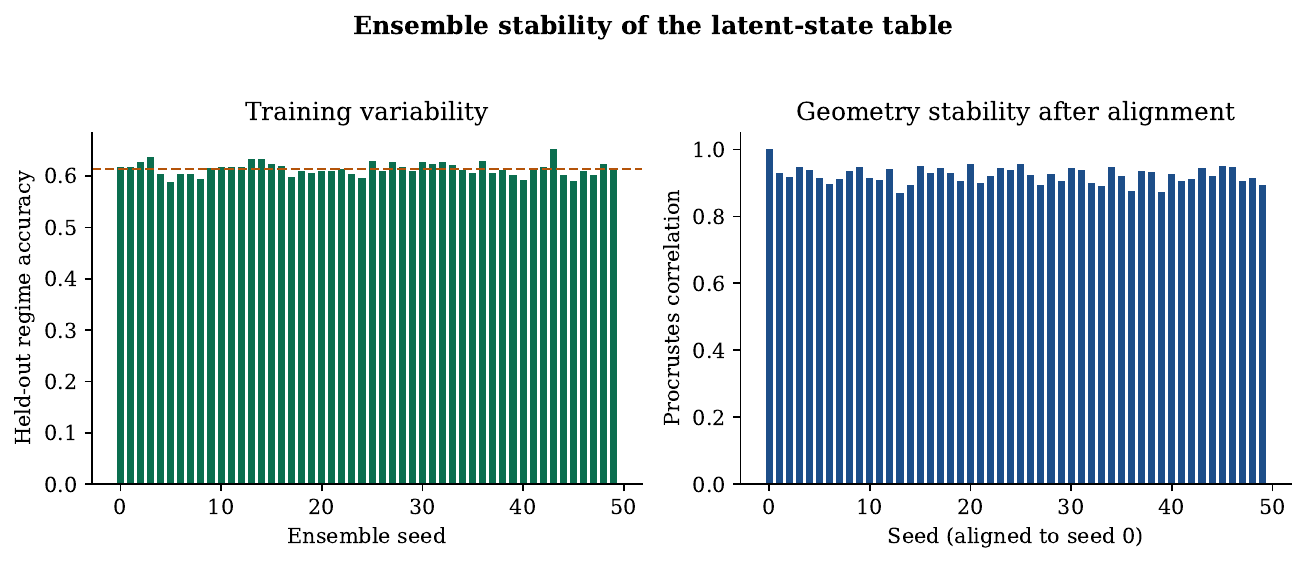}
\caption{Initialization stability: held-out accuracy across seeds and Procrustes correlation of aligned state tables (\(B=50\)).}
\label{fig:unc-seed}
\end{figure}

\paragraph{Uncertainty of the Latent Memory Table (\(P_5\)).}
\label{sec:uncertainty}
The Latent Memory Table is a statistical estimate rather than a deterministic embedding.
Because coordinates are rotation non-identifiable (Remark~\ref{rem:rotate}), we retrain an ensemble of \(B=50\) encoders (fixed split; seeds vary), align replicates by orthogonal Procrustes analysis
\citep{gower1975,schonemann1966},
and report \(\bar m_i(t)\), \(\widehat\Sigma_i(t)\) and row-wise reliability.
Table~\ref{tab:unc} and Figure~\ref{fig:unc-seed} summarize initialization stability; additional diagnostics (coverage association, trajectory bands, heatmaps) are in Appendix~\ref{app:uncertainty}.
Row-wise \(\widehat\Sigma_i(t)\) supports treating \(\widehat m_i(t)\) as a noisy covariate in models fitted on the Latent Memory Table.
Sensitivity to \(d\) is in Appendix~\ref{app:repro}; a fuller grid over \(W\), stride, mask and encoder class remains open.

\begin{table}[t]
\centering
\small
\caption{Ensemble uncertainty for the Latent Memory Table (\(B=50\) seeds; fixed player split; Procrustes-aligned).}
\label{tab:unc}
\begin{tabular}{@{}lc@{}}
\toprule
Quantity & Value \\
\midrule
Test accuracy (mean \(\pm\) SD) & \(0.614\pm 0.013\) \\
Mean Procrustes correlation (to seed 0) & 0.922 \\
Median ensemble cosine similarity & 0.968 \\
Median \(\operatorname{tr}(\widehat\Sigma_i(t))\) & 2.18 \\
Corr\(\bigl(\operatorname{tr}(\widehat\Sigma),\,\mathrm{coverage}\bigr)\) & 0.563 \\
Corr\(\bigl(\operatorname{tr}(\widehat\Sigma),\,\mathrm{mask\ sens.}\bigr)\) & 0.527 \\
\bottomrule
\end{tabular}
\end{table}

%==============================================================================
\section{Simulation study}
\label{sec:sim}
%==============================================================================

The empirical case study in Section~\ref{sec:data} cannot by itself establish that the procedure recovers a latent memory state: real-world \(m_i(t)\) is unobserved.
Simulations supply the complementary half of the argument.
With a known data-generating process we ask whether tables constructed by the same pipeline recover properties of the true memory mechanism under \(\mathcal{P}\) and \(Q\), and whether they fail when no stable memory is present.
The intended message is methodological: \emph{the framework works because it recovers known properties under distinct memory mechanisms}, not because it succeeds on a single observational panel.

\paragraph{Purpose and evaluation geometry.}
\label{sec:sim-purpose}
The design targets settings where the true memory is multivariate, personalized or temporally richer than fixed scalar smoothers.
We do \emph{not} optimize for regime classification accuracy as a success criterion.
Because latent coordinates are rotation non-identifiable, recovery is assessed via Procrustes correlation
\citep{gower1975},
canonical correlations, pairwise-distance correlation and personalization/persistence functionals (Remark~\ref{rem:rotate}), on the \emph{same} \(\mathcal{P}/Q\) scale used for SoccerMon.

\paragraph{Simulation design.}
\label{sec:sim-design}
Four blocks prioritize distinct mechanisms (full factorial deferred to larger Monte Carlo):

\begin{enumerate}[leftmargin=1.4em,itemsep=0.25em]
\item Basic recovery (I):
True lag weights short-exponential, delayed/gamma or biphasic; compare oracle states, Transformer and linear encoders, PCA-lagged windows and classical EWMA-type summaries.
\item Personalization versus regime (II):
Vary subject-specific memory dispersion \(\sigma_\tau\) and whether regimes shift exposures only or also memory parameters \(\psi_i\).
\item Window misspecification (IV):
True horizon \(W_0\in\{14,28\}\) crossed with fitted \(W\in\{7,14,28,42\}\), asking whether \(Q\) and recovery peak near the true memory length.
\item Negative control (V):
No stable memory of exposures (scenario~F): the ladder must not manufacture strong state recovery or interpretability merely from successful regime classification.
\end{enumerate}

The DGP generates \(p\) AR(1) exposure channels with regime means and subject effects, then forms the oracle memory
\[
m^{0}_{ik}(t)
=
\sum_{j=1}^{p}\sum_{\ell=0}^{W_0-1} w_{kj}(\ell;\psi_i)\, x_{ij,t-\ell}.
\]
Estimators see only masked windows of \(x\); they never observe \(m^0\) except for oracle scoring.
Protocol, pseudocode and code: \texttt{memory\_reps/sim/} (\texttt{PROTOCOL.md}; drivers under \texttt{scenario\_A}--\texttt{scenario\_D}).

\paragraph{Results on the \(\mathcal{P}/Q\) scale.}
\label{sec:sim-report}
Pilot Monte Carlo runs (\(N=40\), \(T=90\), three replicates; \texttt{outputs/sim\_nuclear/}) are scored exactly as empirical tables.
Table~\ref{tab:sim-Q} summarizes mean \(Q\) and selected property scores by block and estimator.

\begin{table}[t]
\centering
\small
\caption{Simulation study (pilot Monte Carlo): mean \(Q\) and selected scores by block (three replicates).
Oracle sees \(m^0\); classical~=~ EWMA-type summaries; PCA~=~ lagged PCA.}
\label{tab:sim-Q}
\begin{tabular}{@{}llcccc@{}}
\toprule
Block & Estimator & \(Q\) & \(S_{\mathrm{str}}\) & \(S_{\mathrm{interp}}\) & Procrustes \\
\midrule
I recovery
  & Oracle & 0.71 & 0.11 & 1.00 & 1.00 \\
  & Transformer & 0.66 & 0.89 & 0.57 & 0.39 \\
  & Classical & 0.66 & 0.94 & 0.55 & 0.46 \\
  & PCA-lagged & 0.64 & 0.95 & 0.51 & 0.47 \\
\addlinespace
II personalization
  & Oracle & 0.72 & 0.03 & 1.00 & 1.00 \\
  & Transformer & 0.59 & 0.77 & 0.56 & 0.40 \\
  & Classical & 0.61 & 0.81 & 0.57 & 0.47 \\
\addlinespace
V negative control
  & Transformer & 0.54 & 0.82 & 0.23 & 0.15 \\
  & Classical & 0.57 & 0.86 & 0.30 & 0.16 \\
  & Oracle & 0.39 & 0.02 & 1.00$^{*}$ & 1.00$^{*}$ \\
\bottomrule
\end{tabular}\\[0.25em]
{\footnotesize $^{*}$Oracle \(m^0\) is undefined under no memory; scores reflect the degenerate target used in scoring code.}
\end{table}

\begin{figure}[t]
\centering
\includegraphics[width=0.975\textwidth]{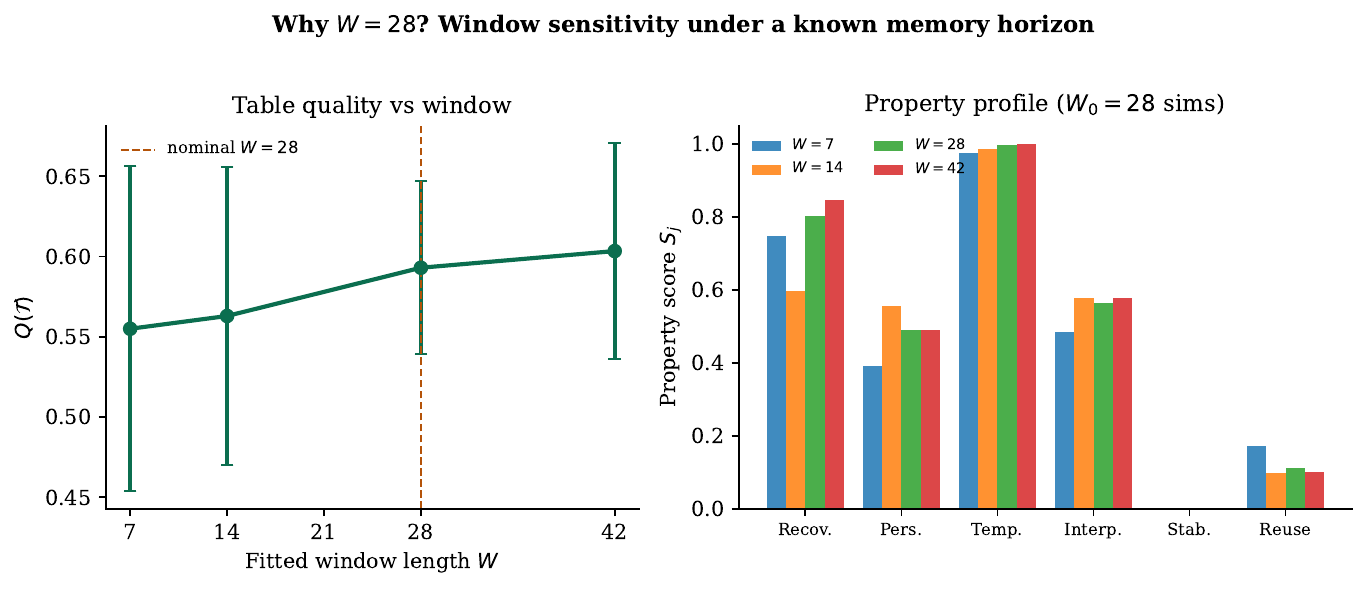}
\caption{Simulation window sensitivity: quality profile versus fitted \(W\) under known true memory horizon (pilot Monte Carlo).}
\label{fig:window}
\end{figure}

Three patterns matter for the methodological claim.
First, under genuine memory (blocks I--II), learned and classical tables both retain regime structure, but Procrustes recovery of \(m^0\) remains imperfect for all non-oracle estimators: the simulation diagnoses \emph{partial} recovery of a known operator, not perfect reconstruction.
Second, under the negative control (block V), Transformer regime accuracy stays high (\(S_{\mathrm{structure}}\approx 0.82\)) while interpretability/state-recovery collapses (\(S_{\mathrm{interp}}\approx 0.23\), Procrustes \(\approx 0.15\)), so \(Q\approx 0.54\).
\emph{Structure alone does not certify a memory table}---precisely the distinction \(\mathcal{P}\) was built to enforce.
Third, under exponential memory with true horizon \(W_0=28\), Transformer \(Q\) peaks at fitted \(W=28\) (\(0.61\)) relative to \(W=7\) (\(0.55\)) and remains comparable at \(W=42\) (\(0.60\))
(Figure~\ref{fig:window}; Table~\ref{tab:sim-W}).
Window length is therefore an empirical property of \(\mathcal{T}\) under \(\mathcal{P}\), not a fixed convention carried over from a domain dashboard or scalar smoother.

\begin{table}[t]
\centering
\small
\caption{Window misspecification (block IV): Transformer \(Q\) when true horizon \(W_0=28\).}
\label{tab:sim-W}
\begin{tabular}{@{}lcccc@{}}
\toprule
Fitted \(W\) & 7 & 14 & 28 & 42 \\
\midrule
\(Q\) (mean) & 0.55 & 0.56 & 0.61 & 0.60 \\
\bottomrule
\end{tabular}
\end{table}

Taken together, the simulations locate success on \emph{mechanism recovery}: discrimination of memory from non-memory, sensitivity to horizon misspecification, and partial reconstruction of multivariate lag structure under the same property system used on SoccerMon.
We also executed a larger Monte Carlo (MC=200) to confirm the qualitative patterns above. For readability, the main text reports the pilot Monte Carlo summaries; full MC summaries and reproducibility materials are provided in the repository.

%==============================================================================
\section{Software architecture and reproducible workflow}
\label{sec:software}
%==============================================================================

The provided codebase is structured around the exported Latent Memory Table, rather than network parameters.
Python is used to build windowed histories, estimate \(\widehat{\mathcal{M}}_\theta\), and export
\(\mathcal{T} = \{\, (\widehat m_i(t),\, \widehat\Sigma_i(t),\, R_i(t)) \,\}\)
as Parquet/Arrow files. Subsequent analyses---including computation of \(Q(\mathcal{T})\), GAMMs, variance decompositions, and manuscript figures---are carried out in R, eliminating the need to refit the encoder.
That language split matches the scientific split between estimating the operator and analysing the table.
Source code, figure-reproduction scripts and reproducibility instructions are available at
\url{https://github.com/idaejin/latent-memory-states}
(Appendix~\ref{app:software}).

%==============================================================================
\section{Reusability as statistical operators on \(\mathcal{T}\)}
\label{sec:applied}
%==============================================================================

Property \(P_6\) (Definition~\ref{def:P}) requires that a fixed Latent Memory Table retain incremental information for targets not used to train \(\mathcal{M}_\theta\).
Once \(\mathcal{T}\) is \(\mathcal{P}\)-validated, that requirement is naturally read as a \emph{family of statistical operators} on the Latent Memory Table---operators that freeze the encoder and act on rows \(\widehat m_i(t)\) (and, where available, \(\widehat\Sigma_i(t)\)).
Section~\ref{sec:incremental} already instantiates one such operator (lagged regression / incremental \(R^2\)).
Here we close the methodological frame by listing the complementary operators that \(\mathcal{P}\) licenses without committing to a domain endpoint.

Formally, write \(\mathcal{A}(\mathcal{T})\) for an analysis map that returns paths, scores, neighbourhoods, partitions or fitted models.
Reusability asks that several \(\mathcal{A}\) remain scientifically meaningful on one exported Latent Memory Table.
The five maps below are not a sports-science menu; they are the standard longitudinal toolkit applied to a moderate-dimensional state table.

The path operator treats \(t\mapsto \widehat m_i(t)\) as a multivariate longitudinal path
(Figure~\ref{fig:traj}).
Summaries include location, displacement, occupation of regions and uncertainty bands from \(\widehat\Sigma_i(t)\); the persistence profile under \(P_3\) is the population counterpart.
This is the functional-data view of \(\mathcal{T}\).

\begin{figure}[t]
\centering
\includegraphics[width=0.975\textwidth]{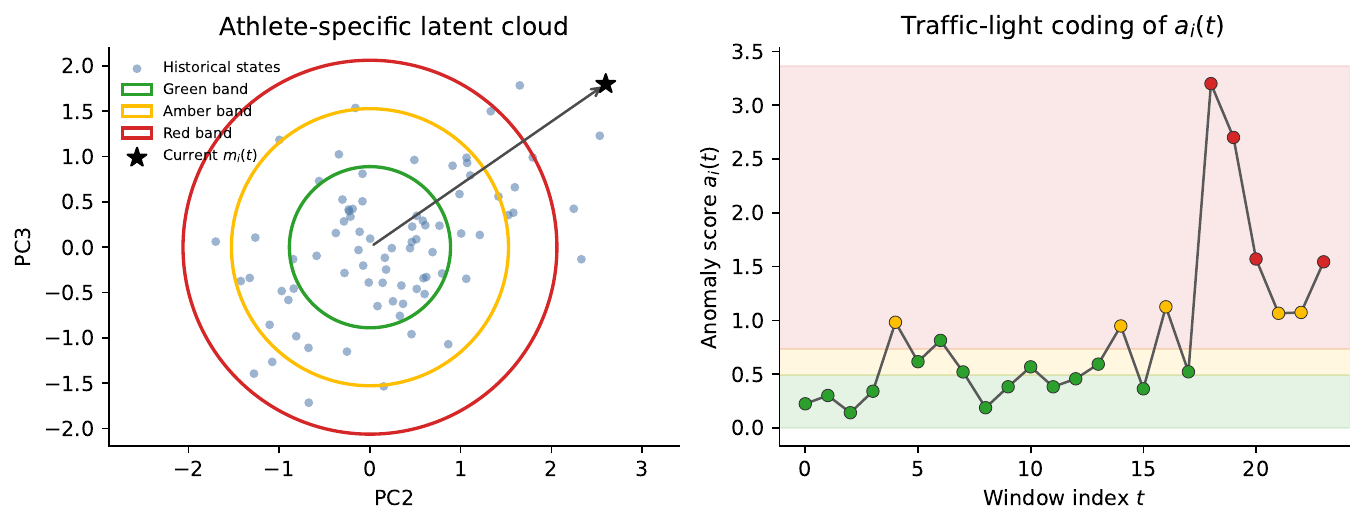}
\caption{Schematic anomaly operator \(a_i(t)\) with quantile coding of the score's own distribution.}
\label{fig:anomaly}
\end{figure}

The anomaly operator, relative to a reference set of past states for unit \(i\),
\begin{equation}
\label{eq:anomaly}
a_i(t)
=
\bigl(\widehat m_i(t)-\widehat\mu_i\bigr)^\top
\widehat\Gamma_i^{+}
\bigl(\widehat m_i(t)-\widehat\mu_i\bigr)
\end{equation}
is a Mahalanobis (or ridge/robust) unusualness score in the latent space
(Figure~\ref{fig:anomaly}).
Quantile coding of \(\{a_i(s)\}\) is a display of the score's empirical distribution, not a clinical threshold; low reliability \(R_i(t)\) should down-weight \(a_i(t)\).

The neighbourhood operator uses \(k\)-nearest neighbours of \(\widehat m_i(t)\) under cosine or Mahalanobis similarity to implement case-based retrieval in the validated metric
(Remark~\ref{rem:rotate}).
Attached covariates and short-horizon successors support nonparametric inspection without a new supervised model per query.

The partition operator clusters rows or path summaries and thereby induces a discrete structure on \(\mathcal{T}\).
Standard unsupervised diagnostics apply; the encoder is not retrained.
Partitions may later enter regressions as discrete covariates, bridging exploratory and confirmatory reuse.

The covariate operator lets rows of \(\mathcal{T}\) enter mixed-effects and other regression models, survival or joint models
\citep{tsiatis2004,rizopoulos2012},
FDA/FPCA or lagged regressions as time-varying predictors, optionally alongside classical fixed summaries \(Z\).
Nested comparisons of \(Z\) versus \(Z\) plus \(\widehat m\) are the empirical content of \(P_6\) already reported for SoccerMon; the same operator class applies whenever \(\mathcal{T}\) is exported.

Table~\ref{tab:applied} records the five operators and their links into \(\mathcal{P}\).
In this reading, Section~\ref{sec:applied} is not an appendix of applications: it is the operational content of reusability.

\begin{table}[t]
\centering
\small
\caption{Statistical operators on a \(\mathcal{P}\)-validated Latent Memory Table (instantiations of \(P_6\)).}
\label{tab:applied}
\begin{tabular}{@{}
  >{\raggedright\arraybackslash}p{0.28\linewidth}
  >{\raggedright\arraybackslash}p{0.42\linewidth}
  >{\raggedright\arraybackslash}p{0.22\linewidth}
  @{}}
\toprule
\textbf{Operator} & \textbf{Map on \(\mathcal{T}\)} & \textbf{Link} \\
\midrule
Path
  & \(t\mapsto \widehat m_i(t)\); projections; uncertainty bands
  & \(P_3\), \(P_5\), \(P_6\) \\
\addlinespace
Anomaly
  & Mahalanobis distance to own history; quantile coding
  & \(P_2\), \(P_5\), \(P_6\) \\
\addlinespace
Neighbourhood
  & \(k\)-NN under cosine / Mahalanobis similarity
  & \(P_2\), \(P_3\), \(P_6\) \\
\addlinespace
Partition
  & Clustering of rows or path summaries
  & \(P_1\), \(P_2\), \(P_6\) \\
\addlinespace
Covariate
  & Mixed, survival/joint, FDA/FPCA, lagged regression
  & \(P_6\) (primary) \\
\bottomrule
\end{tabular}
\end{table}

%==============================================================================
\section{Discussion}
\label{sec:discuss}
%==============================================================================

This paper is not primarily about Transformers or athlete monitoring.
That opening already states the thesis: the Latent Memory Table is the unit of analysis; the encoder, \(\mathcal{P}\), the simulations and SoccerMon exist only to argue that it deserves that status.
Once \(\mathcal{T}\) is \(\mathcal{P}\)-validated, ordinary statistical operators act on it without retraining (Section~\ref{sec:applied}).

Empirical and simulation results support that claim rather than redefine it.
Relative to \(\mathcal{P}\), the simulation study discriminates genuine memory from negative controls and shows window-sensitive \(Q\) under known horizons; Procrustes recovery of \(m^0\) remains partial for non-oracle estimators.
On SoccerMon, the constructed Latent Memory Table is recoverable for Team\(\times\)Season (\(P_1\)), temporally coherent over multi-week lags (\(P_3\)), aligned with load/wellness constructs after regime conditioning (\(P_4\)), and partially informative beyond classical summaries for some next-window targets (\(P_6\)).
Personalization beyond regime (\(P_2\)) is modest under the present auxiliary criterion, though subject-identity classification and some PC variance shares retain unit-level signal.
The Procrustes ensemble supplies row-wise reliability for \(\widehat\Sigma_i(t)\) (\(P_5\) under initialization).
The delivered product remains the table \(\mathcal{T}\) (unit, time, \(\widehat m_1,\dots,\widehat m_d\), uncertainty), admissible later as input to mixed-effects models, functional data analysis, survival and joint models, clustering or FPCA.
This paper does not fit those endpoint models as primary analyses.

A natural objection is that one might simply train an end-to-end predictor of a favoured clinical or operational label.
We reject that as the primary scientific strategy for three reasons.
First, endpoint definitions are often contested and setting-specific; optimizing a table for one label locks the analysis to that label.
Second, outcome-specific supervision forces the encoder to discard information irrelevant to the chosen endpoint but potentially relevant to others.
Third, a fixed Latent Memory Table preserves multi-purpose utility without retraining for each endpoint
\citep{tsiatis2004,rizopoulos2012}.
In short, \(\mathcal{T}\) is a unit of analysis for longitudinal memory---not a task-specific feature map or a disposable embedding; Section~\ref{sec:applied} closes the frame by reading \(P_6\) as operators on the Latent Memory Table.

Limitations remain.
Team\(\times\)Season labels mix coaching, calendars and measurement systems; within-regime partial correlations mitigate but do not eliminate this regime confounding.
A large fraction of SoccerMon channels are self-reports (RPE, wellness), so reporting bias and scale coarseness limit physiological claims even when GAMM smooths are significant (Remark~\ref{rem:subjective}).
Robustness to LSTM / linear baselines remains future work (Appendix~\ref{app:repro} reports sensitivity to latent dimension), and regime classification is only a minimal auxiliary criterion---contrastive or reconstruction objectives may improve personalization.
The empirical scale is modest (two teams, 66 athletes with windows), so external validity is untested.
Finally, masks may encode device protocols versus subjective non-response; disentangling informative versus protocol missingness remains open.

Taken together, classical fixed scalar summaries compress recent exposure into engineered univariate memories, whereas the present framework estimates, validates and exports the Latent Memory Table.
Simulations locate validity in recovery of known memory mechanisms under \(\mathcal{P}\); SoccerMon illustrates the same pipeline in one monitoring domain.
Neither the architecture nor the sport is the thesis: the thesis is that the Latent Memory Table deserves to be treated as a reusable statistical object.

%==============================================================================
\section{Conclusion}
\label{sec:conclusion}
%==============================================================================

The Latent Memory Table (Figure~\ref{fig:product}) is the contribution; the encoder, the simulations and SoccerMon exist to argue that it deserves that status.
We validate \(\mathcal{T}\) under the property system \(\mathcal{P}\) with composite quality \(Q\).
Classical fixed scalar summaries are degenerate special cases of the same operator class; statistical operators on \(\mathcal{T}\) (Section~\ref{sec:applied}) are licensed without retraining once \(\mathcal{P}\)-validity holds.

\section*{Acknowledgements}
This paper was partially funded by the SPHERES project (PID2023-153222OB-I00), granted by MCIU/AEI/10.13039/501100011033 and by ``FEDER, UE''.
This is a preprint. Code and reproducibility materials are archived at
\url{https://github.com/idaejin/latent-memory-states}.

\section*{Declaration of generative AI and AI-assisted technologies in the manuscript preparation process}
During the preparation of this work, the author used Cursor (an AI-assisted coding and writing environment) for drafting and editing manuscript text, assisting with analysis and simulation code, and refining presentation.
The author reviewed and edited the output as needed and takes full responsibility for the content of the published article.

\section*{Data availability}
Analysis code, manuscript sources and reproducibility instructions are archived at
\url{https://github.com/idaejin/latent-memory-states}
(Section~\ref{sec:software}).
Anonymized Latent Memory Tables (Parquet/Arrow exports with states, uncertainty and reliability summaries) can be regenerated with that code or obtained from the authors upon reasonable request.
Raw SoccerMon sensor streams are not redistributed here; the public SoccerMon release is described by
Midoglu et al.\ (2024), \emph{Scientific Data}, \url{https://doi.org/10.1038/s41597-024-03386-x}
(Zenodo: \url{https://doi.org/10.5281/zenodo.10033832}).

%==============================================================================

\appendix

\section{Encoder hyperparameters}
\label{app:arch}

\begin{table}[h]
\centering
\caption{Encoder hyperparameters used in the SoccerMon study.}
\label{tab:arch}
\begin{tabular}{@{}lc@{}}
\toprule
Hyperparameter & Value \\
\midrule
Window length \(W\) & 28 days \\
Stride & 7 days \\
Input channels (values + masks) & \(12+12\) \\
Latent dimension \(d\) & 32 \\
Layers / heads & 2 / 4 \\
Feed-forward width & 64 \\
Dropout & 0.1 \\
Optimizer & AdamW (\(10^{-3}\), weight decay \(10^{-4}\)) \\
Epochs (selected by held-out accuracy) & 20 \\
\bottomrule
\end{tabular}
\end{table}

\section{Missingness coverage}
\label{app:missing}

\begin{figure}[h]
\centering
\includegraphics[width=0.95\textwidth]{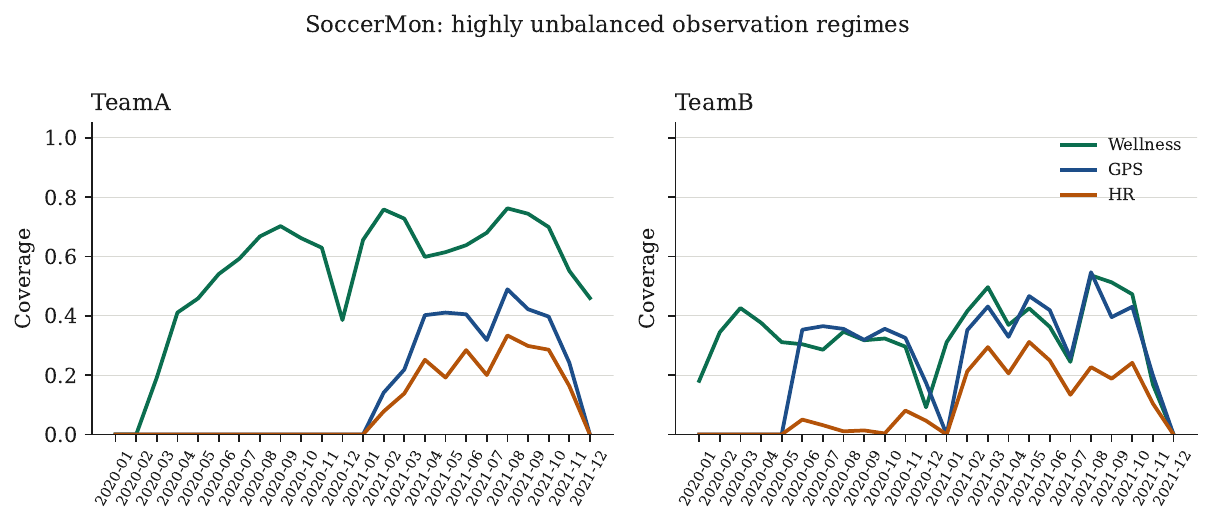}
\caption{Monthly observation coverage for subjective wellness versus objective GPS and HR by team.
Unbalanced regimes and source-specific missingness motivate explicit masks in \(H_{i,t}\).}
\label{fig:missing}
\end{figure}

\section{GAMM specification and functional PCA}
\label{app:gamm}

\begin{figure}[h]
\centering
\includegraphics[width=0.95\textwidth]{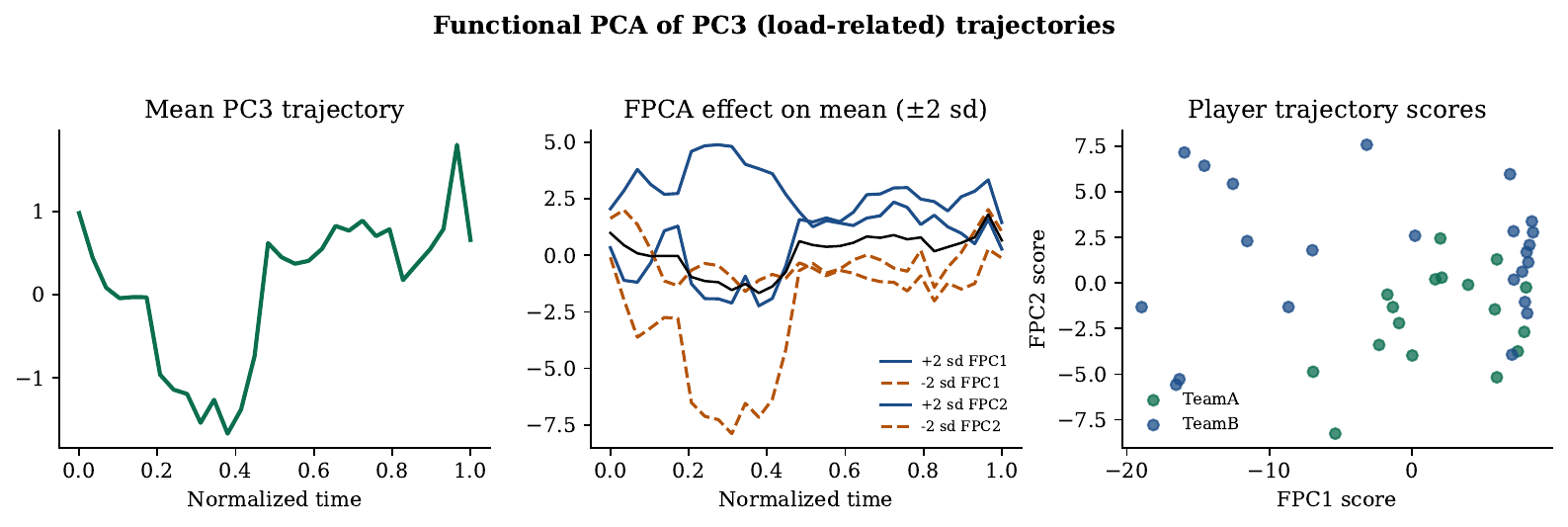}
\caption{Functional PCA of PC3 (load-related) trajectories: mean curve, \(\pm 2\)\,sd FPC effects and player scores.}
\label{fig:fpca}
\end{figure}

We fit GAMMs with the R package \texttt{mgcv}
\citep{wood2017}:
\[
\mathrm{PC}_{k,it}
=
s(x_{it})
+
\alpha_{g(i,t)}
+
b_i
+
\varepsilon_{it},
\qquad
b_i\sim N(0,\sigma_b^2),
\]
where \(g(i,t)\) indexes the Team\(\times\)Season regime of window \((i,t)\) (four fixed-effect levels), \(s\) is a thin-plate smooth with basis dimension \(k=8\), and \(b_i\) is a player random intercept.
Model \(R^2\) values in Table~\ref{tab:gamm} include regime and player terms; partial-effect curves isolate \(s(x)\).
Discrete functional PCA of PC3 trajectories for athletes with dense coverage
(Figure~\ref{fig:fpca})
yields a leading FPC explaining \(\approx 61\%\) of between-player curve variability.

\section{Additional uncertainty diagnostics}
\label{app:uncertainty}

\begin{figure}[h]
\centering
\includegraphics[width=0.95\textwidth]{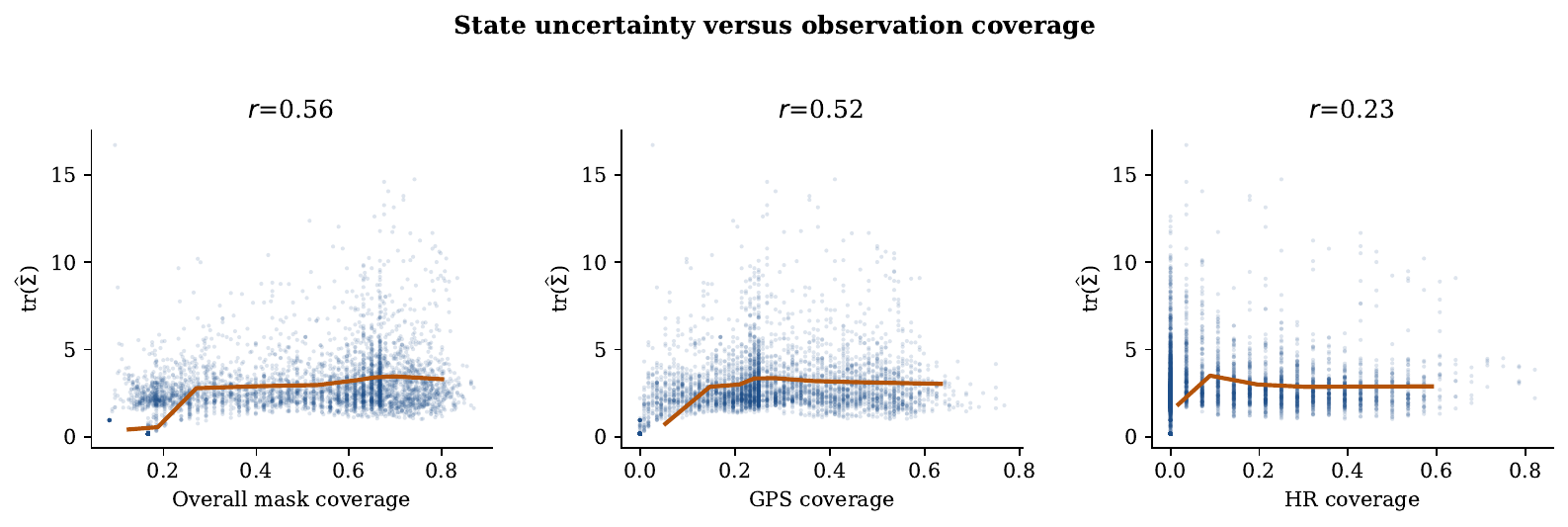}
\caption{Row-wise state uncertainty versus observation coverage.}
\label{fig:unc-cov}
\end{figure}

\begin{figure}[h]
\centering
\includegraphics[width=0.92\textwidth]{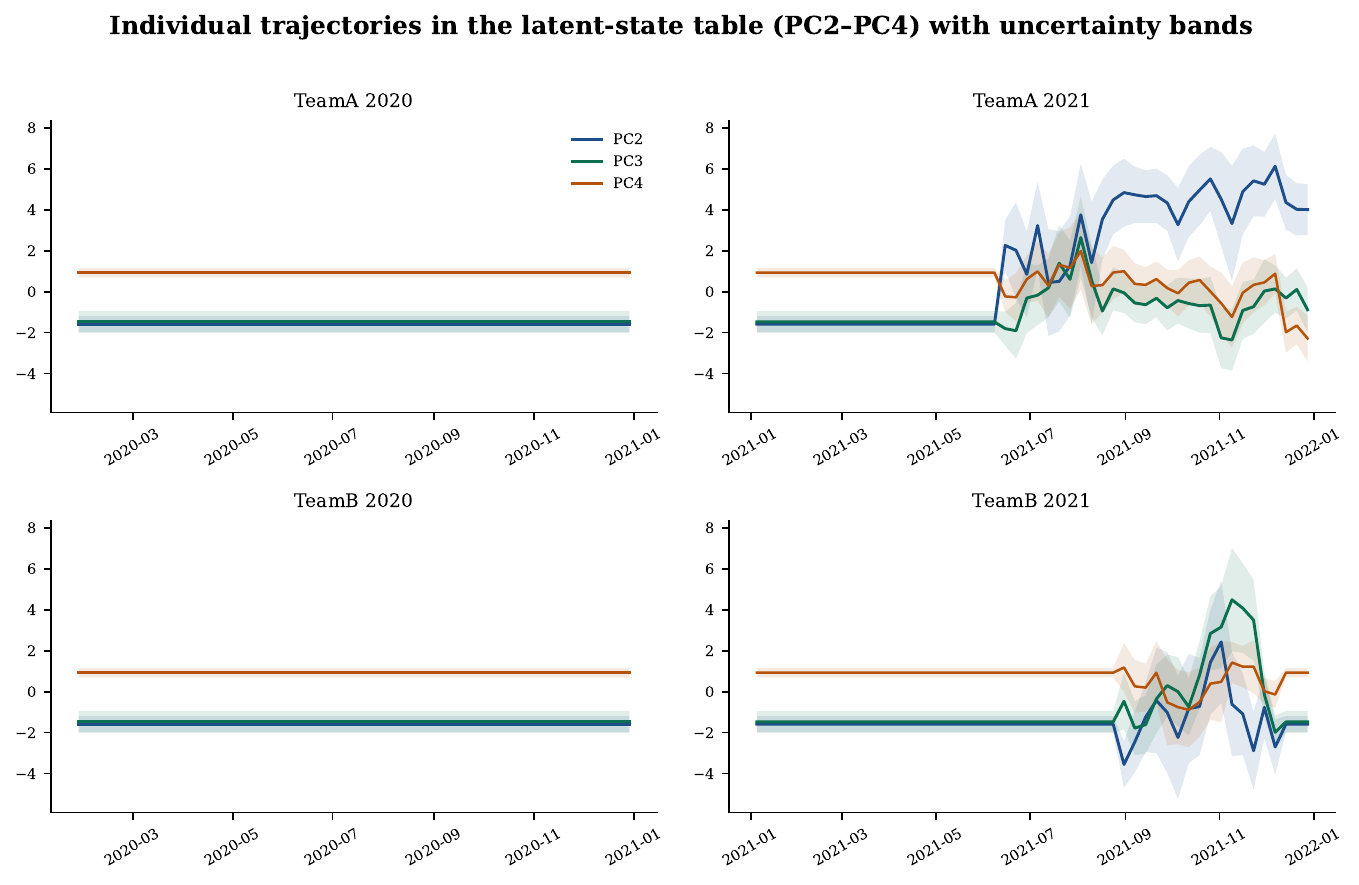}
\caption{Aligned PC trajectories with ensemble bands (\(\bar m\pm 1.96\,\mathrm{SD}\) across \(B=50\) seeds).}
\label{fig:unc-traj}
\end{figure}

\begin{figure}[h]
\centering
\includegraphics[width=\textwidth]{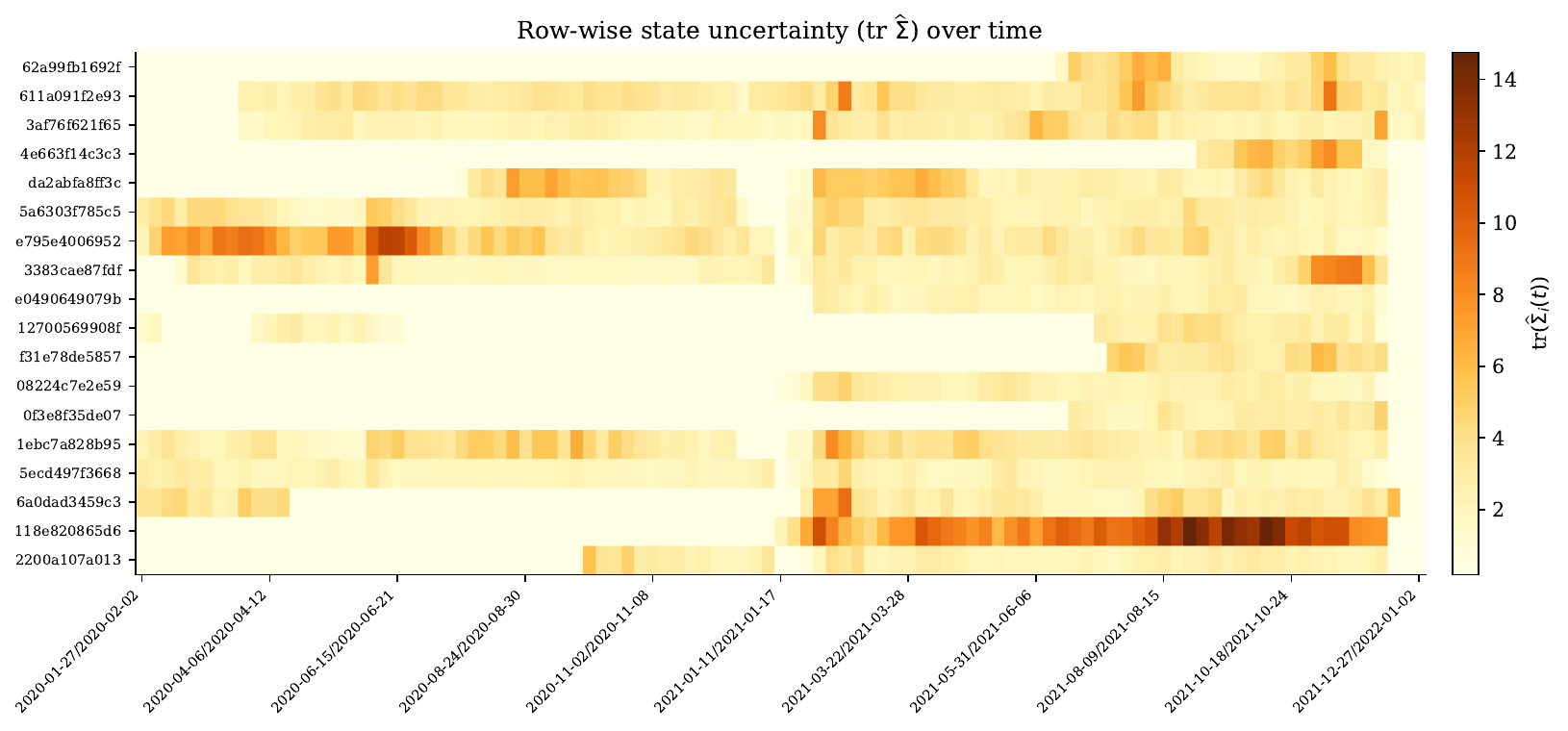}
\caption{Heatmap of \(\operatorname{tr}(\widehat\Sigma_i(t))\) over weeks for athletes with dense coverage (\(B=50\) ensemble).}
\label{fig:unc-heat}
\end{figure}

With the \(B=50\) Procrustes ensemble, disagreement correlates positively with mask coverage (\(r\approx 0.56\)) and mask-perturbation sensitivity (\(r\approx 0.53\))
(Figure~\ref{fig:unc-cov}): sparse windows tend toward a more reproducible default encoding, whereas richly observed windows expose greater seed-to-seed variability.
This is reliability of the estimator, not physiological trustworthiness of sparse rows.
Trajectory bands
(Figure~\ref{fig:unc-traj})
and a player--time heatmap of \(\operatorname{tr}(\widehat\Sigma)\)
(Figure~\ref{fig:unc-heat})
make row-wise reliability visible for use of the Latent Memory Table.

\section{Software details}
\label{app:software}

Code and reproducibility notes are at
\url{https://github.com/idaejin/latent-memory-states}.
The intended pipeline is
raw histories \(\to\) windows \(\to\) operator estimation \(\to\) \(\mathcal{T}\) \(\to\) validation \(\to\) analysis.
Python handles estimation and export of the Latent Memory Table; R handles mixed models, \texttt{mgcv}, FPCA, uncertainty summaries and graphics.
Conceptual \texttt{quality()} returns the six \(\mathcal{P}\) scores and \(Q(\mathcal{T})\); \texttt{plot()} produces diagnostics of \(\mathcal{T}\).
Manuscript figures that diagnose \(\mathcal{T}\) are regenerated from the exported table and ensemble replicates.
Primary scripts live under \texttt{memory\_reps/}; see \texttt{docs/REPRODUCIBILITY.md} in the repository.

\section{Reproducibility and sensitivity}
\label{app:repro}

\begin{figure}[h]
\centering
\includegraphics[width=0.85\textwidth]{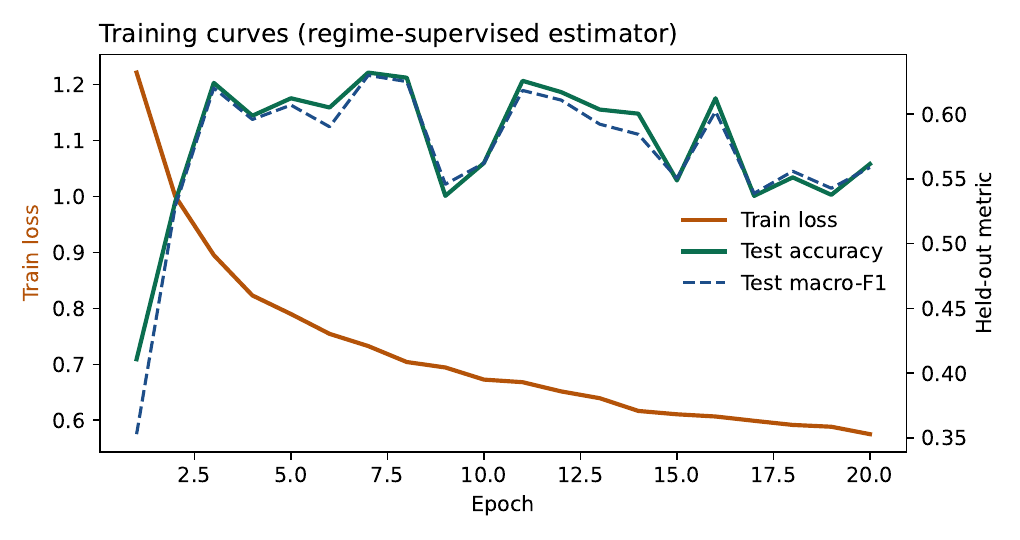}
\caption{Training curves for the regime-supervised estimator (\(d=32\)).}
\label{fig:train}
\end{figure}

\begin{figure}[h]
\centering
\includegraphics[width=0.72\textwidth]{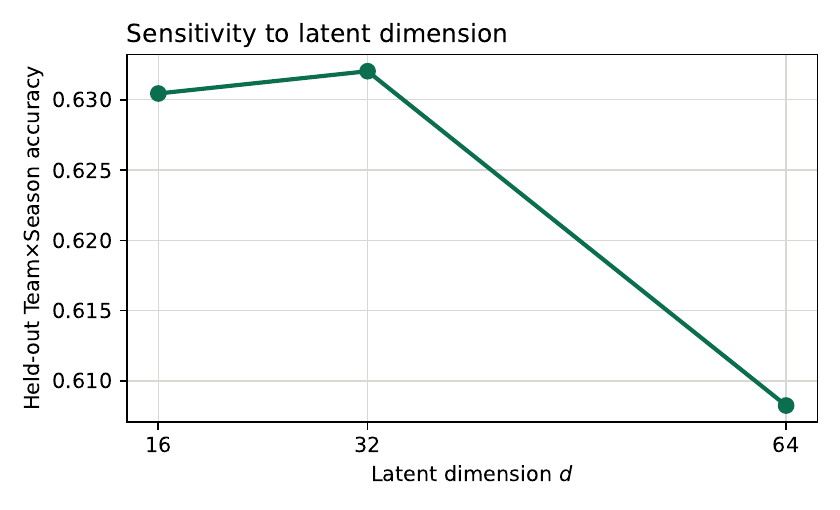}
\caption{Held-out Team\(\times\)Season accuracy versus latent dimension \(d\).}
\label{fig:dim}
\end{figure}

Channel standardization uses training athletes only (Remark~\ref{rem:std}).
Figure~\ref{fig:train} shows training loss and held-out regime accuracy/macro-F1 over 20 epochs for the main estimator (\(d=32\)).
Figure~\ref{fig:dim} and Table~\ref{tab:dim} report sensitivity to latent dimension under the same split and train-only scaling: \(d=16\) and \(d=32\) perform similarly, while \(d=64\) is slightly worse on this panel.

\begin{table}[h]
\centering
\caption{Sensitivity to latent dimension (player-grouped holdout; train-only standardization).}
\label{tab:dim}
\begin{tabular}{@{}cc@{}}
\toprule
\(d\) & Held-out accuracy \\
\midrule
16 & 0.630 \\
32 & 0.632 \\
64 & 0.608 \\
\bottomrule
\end{tabular}
\end{table}

\end{document}